\documentclass[sigconf,screen]{acmart}
\AtBeginDocument{%
  \providecommand\BibTeX{{%
    \normalfont B\kern-0.5em{\scshape i\kern-0.25em b}\kern-0.8em\TeX}}}

\setcopyright{rightsretained}
\acmDOI{10.1145/3664646.3664760}
\acmYear{2024}
\copyrightyear{2024}
\acmSubmissionID{fsews24aiwaremain-p15-p}
\acmISBN{979-8-4007-0685-1/24/07}
\acmConference[AIware '24]{Proceedings of the 1st ACM International Conference on AI-Powered Software}{July 15--16, 2024}{Porto de Galinhas, Brazil}
\acmBooktitle{Proceedings of the 1st ACM International Conference on AI-Powered Software (AIware '24), July 15--16, 2024, Porto de Galinhas, Brazil}
\usepackage{cleveref} 
\usepackage{multirow}
\usepackage{balance}

\newcommand{\timeperiod}{2 week }
\newcommand{\usercount}{34 }          % TODO: UPDATE FINAL N COMPLETIONS
\newcommand{\completioncount}{74k }   % TODO: UPDATE FINAL N DEVS
\newcommand{\devinteraction}{200k }   % TODO: UPDATE FINAL

\newcommand{\jonberta}{{\textsc{JonBERTa}}}
\newcommand{{\jonbertahead}}{{\textsc{JonBERTa-head}}}
\newcommand{{\jonbertaattn}}{{\textsc{JonBERTa-attn}}}

\newcommand{\m}[1]{\multirow{2}{*}{#1}}
\newcommand{\mc}[1]{\multicolumn{2}{c}{#1}}

\newcommand{\mali}[1]{{\color{black}#1}}
\newcommand{\aral}[1]{{\color{black}#1}}

\newcommand{\pluginname}{{\textsc{Code4Me}}}
\newcommand{\reppackage}{{https://github.com/ar4l/curating-code-completions}}
\newcommand{\publishedmodels}{{https://huggingface.co/collections/AISE-TUDelft/smart-invocation-of-code-completion-66473ddf6fa6cf6e541f750c}}
\acmSubmissionID{fsews24aiwaremain-p15-p}
\begin{document}

\title{
  % Exploring how to Curate Code Completions by \\ 
  % Filtering on Code Context and Telemetry Data 
  A Transformer-Based Approach for Smart Invocation of Automatic Code Completion}

\author{Aral de Moor}
\email{a.d.demoor@tudelft.nl}
\orcid{0009-0003-5105-0518}
\affiliation{%
  \institution{Delft University of Technology}
  \city{Delft}
  \country{Netherlands}
}

\author{Arie van Deursen}
\email{arie.vandeursen@tudelft.nl}
\orcid{0000-0003-4850-3312}
\affiliation{%
  \institution{Delft University of Technology}
  \city{Delft}
  \country{Netherlands}
}

\author{Maliheh Izadi}
\email{m.izadi@tudelft.nl}
\orcid{0000-0001-5093-5523}
\affiliation{%
  \institution{Delft University of Technology}
  \city{Delft}
  \country{Netherlands}
}

\renewcommand{\shortauthors}{de Moor et al.}
% endregion 

\begin{abstract}
% The abstract is a short summary of the work to be presented in the article.
% Make it clear and concise. At most 300 words. For a recommended structure, have a look at the introduction section.
%% 192 WORDS

% Large Language Models (LLMs) are effective for code completion, 
% but little research exists on improving the usability of these tools 
% in their new workflows. 
% Particularly, the question of \textit{when} a 
% developer would actually like to see a completion is neglected.

% mali: I am re-writing the abstract as I go
% Transformer-based language models are effective for code completion, 
% and much research focuses on improving the \textit{content} of completions. 
% However, this neglects the usability-interplay with developers;
% and particularly ignores the question of 
% \textit{when} a developer would like to see a completion. 
\mali{Transformer-based language models are highly effective for code completion, with much research dedicated to enhancing the \textit{content} of these completions. 
Despite their effectiveness, these models come with high operational costs and can be intrusive, especially when they suggest too often and interrupt developers who are concentrating on their work. 
Current research largely overlooks how these models interact with developers in practice and neglects to address \textit{when} a developer should receive completion suggestions.}
% To address this, we explore code completion filters trained 
% to show only those completions a developer would like to see. 
\mali{To tackle this issue, we developed a machine learning model 
that can accurately predict when to invoke a code completion tool 
given the code context and available telemetry data. 
% This approach helps to reduce costs and minimizes interruptions for developers.
}

% We collect a dataset of 200k developer interactions with code completions from a cross-IDE plugin developed at our institution.
%%
% To do so, we collect a dataset of \devinteraction developer interactions 
% with our cross-IDE code completion plugin.
% Using this interaction data, we train invocation filtering models
% to intelligently invoke the completion model only 
% when a developer wants to see a completion.
\mali{To do so, we collect a dataset of \devinteraction developer interactions 
with our cross-IDE code completion plugin and train several invocation filtering models.}
% We find that transformer models perform best at this task 
% while maintaining low enough latency. 
Our results indicate that our small-scale transformer model significantly outperforms the baseline while maintaining low enough latency.
% TODO: considerably outperform a baseline we create by reverse-engineering copilot.
\mali{We further explore the search space for integrating additional telemetry data into a \textit{pre-}trained transformer directly and obtain promising results. } 
% Our online evaluation in a code completion plugin, 
% with \completioncount invocations spanning \usercount developers, 
% shows potential for our proposed, hybrid transformers in practice.
\mali{To further demonstrate our approach's practical potential, 
we deployed the model in an online environment with \usercount developers
and provided real-world insights based on \completioncount actual invocations.}

% \todo{Add results, recommendations, and findings.} 

% incorporating scalar-valued telemetry data into transformer-based classification models.
% Additionally, we are the first study to (openly) evaluate our filters in both an offline and online setting.

% The offline results indicate that our novel transformer models attain the best classification accuracy,
% proving they are able to leverage the joint code context and telemetry feature data. 
% Our online evaluation in a real code completion plugin,
% with \completioncount completions    
% spanning \usercount developers, 
% demonstrates that this also holds in practice. 
% By showing that transformers can leverage multi-modal data in classification tasks, 
% this lays out the groundwork for future studies incorporating telemetry feature data 
% directly into the prediction model itself, to curate also the content of completions to developers' current mode of thought. 

\end{abstract}

% TODO: ADD Software and its engineering -> Integrated and Visual development Environments
% region ACM Tags


% TODO: not sure whether these need to be standardised?

% endregion ACM Tags
\maketitle

\section{Introduction}
\label{sec:introduction}

% The introduction is a crucial part of any paper. A compelling introduction immediately captures the reader's interest and answers the question: ``Why should I read this?" Here is the structure I use for my paper introductions. Feel free to adapt it to your needs or develop your own structure.

%% 1-2 sentences to give context about the work
%% LLMs for SE, code suggestions, developer productivity
% Transformer-based code completion has proven its growing capacity 
% at assisting development~\cite{ziegler_productivity_2022},
% with a vast proportion of developers now using Artificial Intelligence (AI)-enhanced tools to write code~\cite{stack_overflow_stack_2023}.
% Most recent AI-powered code completion systems 
% work by prompting a Large Language Model (LLM) 
% with the code context before the cursor 
% \cite{chen_evaluating_2021,izadi_codefill_2022}.      % Codex paper (original)
Transformer-based code completion 
has become essential in modern software development~\cite{ziegler_productivity_2022}. 
The widespread adoption of Artificial Intelligence (AI) tools 
in coding highlights the significance of these models, 
particularly those built on the Transformer architecture~\cite{stack_overflow_stack_2023}. 
These AI-driven tools typically analyse the code preceding the cursor  
to suggest the next lines of code~\cite{chen_evaluating_2021,izadi_codefill_2022}.
% Nowadays, the context after the cursor~\cite{
%   bavarian_efficient_2022,        % FIM from OpenAI (original)
%   roziere_code_2023,              % CodeLlama using FIM
%   lozhkov_starcoder_2024,         % StarCoder2 using FIM
%   izadi_language_2024,            % code4me
%   guo_deepseek-coder_2024}        % DeepSeek-Coder using FIM
% as well as snippets from related files are sometimes included~\cite{ % Optionally: APIGen
%   lu_reacc_2022,                  % ReACC (RAG for Code)
%   wang_codet5_2023}.              % CodeT5+ retrieves relevant code for its encoder
Recent advancements have expanded 
the considered context to include not only the subsequent code~\cite{bavarian_efficient_2022,roziere_code_2023,lozhkov_starcoder_2024,izadi_language_2024,guo_deepseek-coder_2024},
but also related snippets from other files 
to enrich the prediction accuracy~\cite{lu_reacc_2022, wang_codet5_2023}.
% While retrieval-augmented generation strategies 
% may improve the content of a completion, 
% it neglects the usability-interplay with developers' 
% experience, preferences, and current mode of thought 
% \cite{
%   prather_its_2023,               % novice developers & copilot 
%   vaithilingam_expectation_2022,  % cognitive overhead, difficult completions, lack of trust
%   barke_grounded_2022}.           % explorative & accelerative interactions
%% THE PROBLEM, SPECIFICALLY
% Specifically, 
% \citet{barke_grounded_2022} and \citet{liang_understanding_2023} advocate for streamlining interactions where a developer knows what they want to do, and uses code completions to get there faster. 
% Others also highlight the need for better alignment with developers' workflows
% ~\cite{prather_its_2023,vaithilingam_expectation_2022,wang_investigating_2023};
% however, a suggestion delivered at the wrong time or in a faulty code context can be determinental to their flow. 

\mali{This focus on augmenting 
the content quality of completions 
has inadvertently overshadowed 
a vital aspect of the user experience: 
the interaction dynamics between the developers and the AI tools~\cite{prather_its_2023,vaithilingam_expectation_2022,barke_grounded_2022}.
While these models generate high-quality code suggestions, 
their operational and environmental costs pose significant challenges~\cite{chien2023reducing,chatgpt_cost}.
Moreover, due to their potential to disrupt the coding workflow of developers, 
the frequency and timing of these suggestions is critical 
for the overall productivity the tools aim to boost~\cite{prather_its_2023}.}

%% THE GAP WE ADDRESS
% Existing work aims to address this by training a filter 
% to only invoke the completion model 
% when there is a certain guarantee the completion will be accepted~\cite{mozannar_when_2023,sun_dont_2023}.
% We reformulate the problem to leverage the code-context and telemetry available in an IDE, and propose a novel metric for online evaluation.
% We build on their work and reduce the dependence on the underlying completion-model 
% by considering manually-invoked completions as positive, regardless of whether they are accepted or not. 
% Furthermore, the real-world latency of their solutions remains unaddressed. 

% TODO: uncomment
Previous efforts focus on developing a filtering model 
designed to \textit{show} a completion only when there is a high confidence it will be accepted
\cite{mozannar_when_2023,sun_dont_2023}.
This reduces inference cost and likely improves developers' focus.
However, \citet{sun_dont_2023} assume completions are rejected based on the context before the cursor alone, ignoring the interplay with developers' mode of thought. 
\citet{mozannar_when_2023} improves on this by considering in-IDE telemetry data.
However, they propose a relatively complex ensemble model that can incur additional latency by filtering after a completion is generated; and do not consider that some completions, despite being rejected, may help guide the user in their thinking. 

% Previous efforts focus on developing 
% a filtering model designed 
% to \textit{show} the completion model's generation 
% only when there is high confidence 
% that the already-generated completion will be accepted~\cite{mozannar_when_2023}.
% While this method enhances developers' focus 
% by suggesting fewer but more effective completions, 
% it does not diminish the inference cost. 
% This is because it initially activates LLMs, 
% and then selectively display their outputs.
% \mali{Sun et al.~\cite{sun_dont_2023} 
% also developed a classifier 
% as a quality estimation model for generations.
% However, they assume completions are rejected based on code context alone, ignoring the interplay with developers' mode of thought.}

% our approach
In this study, we take a further step to \textit{proactively} predict 
when to \textit{invoke} a code completion model 
based on code context and telemetry data.
Our lightweight, transformer-based filtering model, \jonberta, 
is designed to trigger a code completion model 
only when there's a strong likelihood 
that a developer requires assistance 
\textit{or} is likely to accept the suggested completion. 
To train our model, 
we leverage the data we have collected 
from developers' real-world interactions 
with our open-source code completion tool 
called \textsc{Code4Me}\footnote{https://code4me.me},
available for both VSCode and Jetbrains IDEs.
% \footnote{This plugin has been previously published along with a study at ICSE.  However, following the double-blind policy, we have anonymised its name and will disclose it upon acceptance.  This study presents a problem definition distinct from prior research, with unpublished data and methodology.}.
% {\pluginname} assists developers by line completion in both VSCode and JetBrains IDEs. 
% It integrates three LLMs: Incoder~\cite{fried2022incoder}, UniXcoder~\cite{guo2022unixcoder}, and CodeGPT~\cite{lu2021codexglue}, trained on various programming languages. 
We gather code context and telemetry data from user interactions 
with the plugin, subject to their consent.
We use two indicators to gauge when a developer 
would prefer to receive a suggestion based on usage data: 
(1) when they accept a model-suggested completion, 
and (2) when they manually invoke the model, 
irrespective of whether they ultimately accept or reject the suggestion.
% Moreover, we believe that when 
% developers manually initiate code completion requests 
% (regardless of them being accepted or rejected), 
% it indicates a need for assistance at that moment. 
% We employ an enhanced transformer-based encoder model to accurately predict the optimal moments for activating an LLM-based code completion feature.

% We build upon \citet{mozannar_when_2023} efforts 
% but reduce reliance on the completion model by treating manually triggered completions as positive signals, 
% even if they are not ultimately accepted. 
% Additionally, we address the issue of real-world delay, which was overlooked in their solutions.

% We do so by using the code-context and telemetry data within an IDE. 
% We also introduce a new metric for measuring performance in the real-time setting.

% We further observe among our \devinteraction interactions, that there are times when users manually invoke the completion model, but end up rejecting the completion. 
% \todo{complement here so that i revise it first thing in the morning}
% previous work \cite{mozannar_when_2023,sun_dont_2023} do not address this

% experiments / RQs / results
% \todo{Aral, write this part so that i can revise it early morning}
\mali{Every keystroke made by a developer provides two key types of contextual information that assist our invocation-filtering model in deciding whether to trigger the LLM-based completion system. 
These are: (1) the coding context surrounding the point of invocation, 
and (2) telemetry data gathered via the plugin, e.g. the time since the last completion.
First, we use code context alone and train a transformer-based classifier.}
% Then, aiming to leverage the telemetry data that is collected in an IDE, 
% we explore hybrid transformer architecture to integrate these additional features, such as the time since the last completion, with their contextual understanding. 
\mali{Next, we investigate hybrid transformer architectures 
integrating telemetry data as additional features.}
% As we find these novel models show promise, 
% we quantitatively evaluate them in a user-study with \usercount developers, 
% to investigate how they fare in practice.
\mali{Based on our promising results, 
we evaluate our approach in a user-study with \usercount developers, 
to investigate how our filters perform in practice.}
To this end, we also propose a new performance metric 
to help mitigate issues from optimising for just acceptance rate in previous work ~\cite{mozannar_when_2023,ziegler_productivity_2022}, 
which weighs proxies for the quality and timing of a completion equally as a harmonic mean.
We find that our proposed {\jonbertahead} model scores highest in both the offline and online evaluation. 

Our contributions are as follows: 
\begin{itemize}
  % \item \mali{Our invocation filtering model and its offline evaluation which demonstrates transformer models can improve classification accuracy considerably over the baseline.}
  \item An offline evaluation of a transformer model we fine-tuned on our collected code completion dataset, demonstrating that code context can considerably improve filtering accuracy over a baseline trained on telemetry features only.
  \item \jonberta, a novel transformer architecture to show the potential of training jointly on code context and telemetry data; as well as a custom tokenisation strategy centred on the cursor position.
  \item An online evaluation of our filters in a code-completion plugin with \completioncount invocations, spanning \usercount users.
  % demonstrating practical feasibility, as well as empirical differences between offline and online evaluations. 
  \item For reproducibility purposes, 
  we publish our replication package\footnote{\reppackage}
  with an online appendix, as well as our fine-tuned models\footnote{\publishedmodels}. However, in compliance with the GDPR, we cannot share our training dataset. 
%   we publicly release our replication 
%   % package~\footnote{https://anonymous.4open.science/r/curating-code-completions}, 
%   package\footnote{\reppackage{}}, with an online appendix. 
%   Moreover, we publish our fine-tuned models\footnote{\publishedmodels}.}
% % TODO: https://huggingface.co/AISE-TUDelft}.
\end{itemize}

\section{Background and Related Work}
\label{sec:related-work}

%% COPILOT & CODE COMPLETION 
Today, the most prominent AI-powered code completion tool with over one million active users is GitHub Copilot~\cite{copilot_2021}. %\footnote{https://github.com/features/copilot}.
This tool is powered by a transformer-based~\cite{vaswani_attention_2023} LLM trained on source code, originally with the objective of predicting a function body given its documentation~\cite{chen_evaluating_2021}. 
However, given that the suffix lines below the function are unavailable to the model, the current model powering Copilot is presumably trained with a Fill-In-the-Middle (FIM) objective~\cite{bavarian_efficient_2022}; 
where the model is trained to predict a span of arbitrary length between the prefix and suffix. 
Several alternatives exist to Copilot, namely 
Amazon CodeWhisperer~\cite{codewhisperer_2023}, %\footnote{https://aws.amazon.com/codewhisperer/},
TabNine~\cite{tabnine_2023}, %\footnote{https://www.tabnine.com/}, 
Codeium~\cite{codeium_2023}, %\footnote{https://codeium.com/}, 
Sourcegraph Cody~\cite{cody_2023}, %\footnote{https://sourcegraph.com/cody},  
JetBrains AI~\cite{jetbrains_ai_2023}, and %\footnote{https://www.jetbrains.com/ai/, which offers interfaces beyond code-completion}, and
Gemini Code Assist~\cite{gemini_code_assist_2023}. %\footnote{https://cloud.google.com/gemini/docs/codeassist/overview}.

%% COPILOT USER STUDIES 
% TODO: Though in this paper, we only compare to Copilot as, to our knowledge, all previous studies are based on Copilot
% TODO: consider tagging each with (paid, closed-source)
Since its inception, several user studies and surveys~\cite{
  barke_grounded_2022,            % we know these two by heart now
  ziegler_productivity_2022,      % 
  vaithilingam_expectation_2022,  % 24 devs try programming tasks with copilot
  prather_its_2023,               % usability & interactions for Novice devs
  mozannar_reading_2023,          % differences in user behaviour with copilot
  peng_impact_2023,               % 50% shorter task completion with Copilot
  wang_investigating_2023,        % communicating trust, and configuring AI tools
  liang_understanding_2023}       % developer survey 
% TODO: and design studies? e.g. mcnutt
have been performed on GitHub Copilot\footnote{
  As of 29 December 2023, GitHub Copilot, and others, also offer a conversational interface; 
  but we limit the scope of this paper to generative code completion only. 
}. %~\cite{chen_evaluating_2021}. 
% retrieved from https://github.blog/2023-12-29-github-copilot-chat-now-generally-available-for-organizations-and-individuals/
They highlight that the transformer models backing such tools 
excel at providing \textit{contextual} suggestions,
% TODO: Can cite prather, liang explicitly again
due to their semantical understanding of code.
As a result, this leads to increased developer satisfaction~\cite{vaithilingam_expectation_2022}
and perceived productivity~\cite{ziegler_productivity_2022}.

\subsection{Code Completion Pain Points}
\label{sec:code-completion-pain-points}

Nonetheless, such new technology comes with new questions about its usability and design, 
arising from developer pain points, such as:
distraction due to the \textit{always-on} nature of suggestions~\cite{prather_its_2023},
out-of-distribution generation leading to hallucinated terms~\cite{johnson_r-u-sure_2023}, % Uncertainty-Aware code suggestions
and a lack of personalisation with developers' mode of thought~\cite{barke_grounded_2022}.
Copilot's authors themselves reported that about two-thirds of \textit{shown} completions are ignored by their end-user
\cite{ziegler_productivity_2022}.
Furthermore, GitClear %\footnote{https://www.gitclear.com/} 
recently released a report empirically describing a strong correlation between the adoption of AI code completion and code churn in industry-grade software engineering 
\cite{harding_coding_2024}, 
raising questions about the impact of AI on software maintainability. 
Other studies further find that the bugs introduced by LLM-powered code completion 
are often more subtle~\cite{prather_its_2023,chen_evaluating_2021}, 
and difficult to revert~\cite{vaithilingam_expectation_2022}. 
% List the highlights of approaches that try to solve this problem.

%% DEVELOPERS AND ALIGNING WITH THEM 
\citet{barke_grounded_2022} find that developer interactions with AI-powered code completion are bimodal: 
either \textit{accelerative},
where the developer knows what they want and uses the tool to get there faster; 
or \textit{explorative}, 
where the developer relies on the tool to suggest possible approaches. \citet{prather_its_2023} observe two additional modes among novices: 
shepherding, where a novice slowly accepts a suggestion; and drifting, where they are led down a cyclic `debugging rabbit hole'. 
Not only is this an inefficient use of computational resources, 
but this also misaligns the tool with developer intent. 

Some studies advocate for allowing users to configure the timing and context of completions~\cite{
  wang_investigating_2023,  % Configure AI by customising timing, characteristics & local context
  barke_grounded_2022} % Control over context
to address these concerns.     
% we assume it is better for these tools to adapt to the user, than the user to adapt to or configure these tools. 
However, we assume that the majority of end-users will likely expect such tools to work out of the box and adapt to their usage patterns. 
The issue of language model alignment 
is as pressing as the rate of their increasing capabilities. 
As LLMs are becoming more integrated into end-user workflows, it is necessary to think beyond aligning mere content; but, also the interactions with this content. 

\subsection{Existing Solutions}
\label{sec:existing-solutions}

\citet{sun_dont_2023} aim to address this problem by
filtering out completions that are likely to be rejected. 
They train a transformer-based classifier on the code prefix before it is submitted to the completion model, and find it can hide 5\% of suggestions that would have been rejected with 94.5\% accuracy. 
However, the authors rely on the assumption that rejected completions are due to insufficient code context alone.

%  an extreme-gradient-boosting model 
\citet{mozannar_when_2023} train a filter consisting of two model ensembles: one ensemble before and one after a completion is generated. 
They cite it can hide 53\% of completions with the guarantee 91\% would have been rejected by the user.
However, despite designing their tool for GitHub Copilot, they do not include a user-study.
Additionally, their filter can depend on a completion being generated, which is guaranteed to incur additional latency and compute in practice. 
Moreover, both of these tools aim to maximise the acceptance rate of the shown completions, 
while the need for better alignment with end-users has been highlighted~\cite{
  barke_grounded_2022,
  mozannar_when_2023,
  vaithilingam_expectation_2022,
  russo_navigating_2023}. 

% copilot's logres 
Furthermore, as detailed in a reverse-engineering blog-post
\cite{thakkar_copilot_2023},
Copilot also has its own logistic-regression classifier 
to filter out completions based on telemetry data. 
% This is a lightweight implementation of \citet{mozannar_when_2023}, which hides 5\% of completions with 99+\% guarantee they would have been rejected [Mozannar, personal communication]. 
We analyse how it weighs each feature in more detail in our online appendix, and design our baseline in this study as similar as possible. 

A head-runner in this field would be Gmail SmartCompose~\cite{chen_gmail_2019}, considering the same problem in a natural-language email setting. The authors detail a deep consideration for latency, scalability, and personalisation. However, their approach uses legacy language model architectures; and their insights, while valuable, may not entirely apply to the programming process.

% TODO maybe: The rapidly-improving capabilities 
%~\cite{energy-study-showing-transformers-would-still-improve-even-if-size-was-kept-constant} 
% of transformer models warrant an equal effort at aligning those capabilities with human users. 

% TODO Potentially: Multimodal transformers 
% (need to look how they actually work...)

\section{Problem Definition}
\label{sec:problem-definition}

\mali{Code completion aims to improve developer productivity 
by saving them keystrokes and keeping them in their flow.}
\citet{ziegler_productivity_2022} 
proposes the acceptance rate of suggestions 
as a proxy for developers' perceived productivity; 
however, this is critiqued by \citet{mozannar_when_2023}, 
% following Goodhart's law: ``when a measure becomes a target, it ceases to be a good measure'', 
whom find that optimising the acceptance rate 
results in shorter and typically less useful suggestions. 
% combining this with what I have read in the aforementioned user studies, 

Therefore, some completions, despite seemingly helpful, 
can potentially be detrimental to software quality and programming flow.
And, conversely, some completions, despite being rejected, 
are not wholly ignored by the developer and are potentially helpful in guiding their thinking 
(e.g., a tip-of-the-tongue function call, but with the wrong arguments). 
Determining the actual quality of a completion, 
based on its functional correctness and human preferences for style, 
remains an open question, though research is progressing in the right direction~\cite{evtikhiev_out_2023,zhou_codebertscore_2023,zhuo_ice-score_2024}.

\begin{figure}[h]
    \centering
    \includegraphics[width=\linewidth]{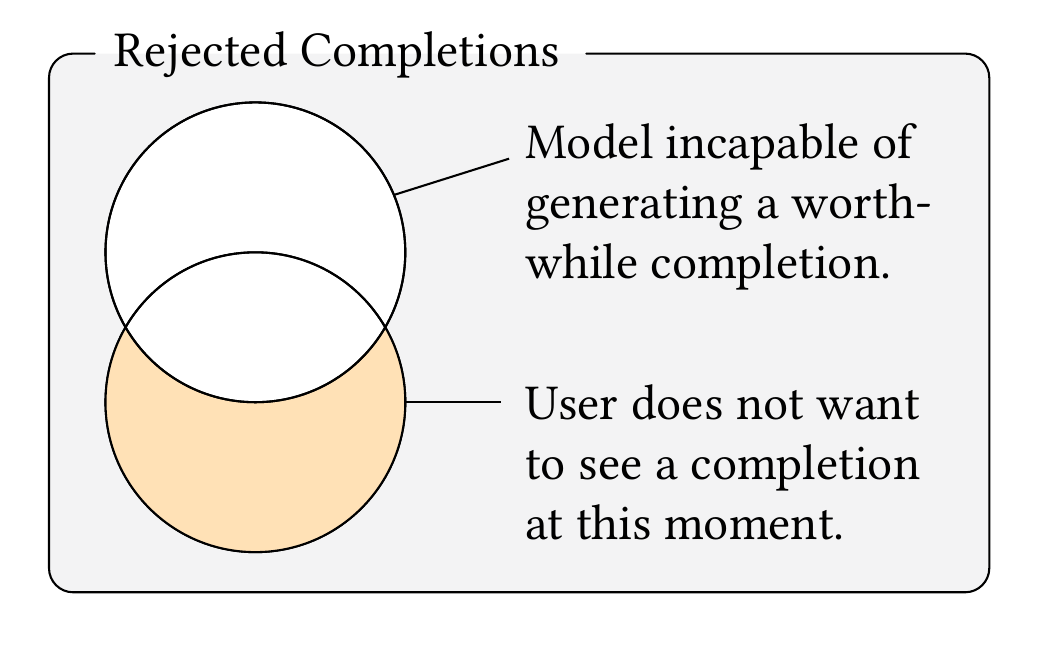}
    \caption{Reasons for Rejected Completions.}
    \label{fig:venn-rejected-completions}
    \Description
      [Venn Diagram of Rejected Completions]
      {Depicts two sets with overlap: 'model incapability', and 'user does not want to see a suggestion'. }
\end{figure}

Consequently, we reframe this problem to what we can actually measure. 
We consider two plausible reasons why a user may have rejected a suggestion, as depicted in 
\Cref{fig:venn-rejected-completions}: 
(1) the model is incapable of generating a good suggestion, which is mainly code context-dependent; or 
(2) the user does not want to see a suggestion, which is mainly user telemetry-dependent. 
These reasons are not exclusive and likely have considerable overlap. 
Thus, we motivate the need for integrating these feature types to better align with the end-user's flow. 

\subsection{Code-Completion Data and Constraints}
\label{sec:problem:code-completion-data-and-constraints}

Throughout this study, we leverage \pluginname, an open-source code completion plugin with around 100 monthly active users developed at our institution 
\cite{izadi_language_2024}. 
% The plugin is available for both JetBrains and VSCode platforms. 
We note three key differences compared with the popular code completion plugins mentioned earlier (\Cref{sec:related-work}): 
% which are relevant for the context of this paper, and the generalisability of our 

\begin{itemize}

  \item \textbf{Line-completion only}. Other tools tend to provide multi-line suggestions in ghost-text style; while we provide completions up to a newline character. Additionally, completions are displayed in a completion box along with typical language-server suggestions. 

  \item \textbf{Restrictive activation}. Contrary to alternative plugins, we provide completions only on a predefined set of trigger characters; at which IDE-based autocomplete would typically trigger (e.g., full stop or an opening parenthesis). Though, we do allow users to manually invoke a completion. 
  % , using the same keybind as the IDE-based autocomplete. 
  
  \item \textbf{Model selection}. Our line-completions are generated by smaller language models than the industry-standard, allowing the user to pick one of three completions provided by 
  \textsc{InCoder}~\cite{fried_incoder_2023},
  \textsc{UniXCoder}~\cite{guo_unixcoder_2022}, and 
  \textsc{CodeGPT}~\cite{lu_codexglue_2021}. 

\end{itemize}

To help keep developers in their flow, we want to leverage language models' ability to complete code even when not at a predefined trigger point. 
In practice, this means that our plugin will query the completions server at any cursor position, and we would like to filter out those points where it may not be necessary to generate a completion. 
Thus, the plugin should filter out those \textit{automatic} suggestions that historically ended up being \textit{rejected} or ignored by the user. 
And, additionally, the plugin infers that a user wants a suggestion, 
at a historically \textit{manual} trigger point. 

As added context, our filter should prioritise false-positives over false-negatives. 
We assume that this is where user preferences lie, given the constant-suggestion nature of the more popular plugins. 
However, to avoid wasted compute and developer distraction, we aim to minimise those completions that are certain to be ignored;
e.g., because the developer is actively typing, or their intent cannot be accurately inferred at the current cursor position. 

Lastly, in terms of non-functional constraints, we would like our filter to take at most 10ms to make a decision. This is in preparation for a backend migration to the 
\textsc{vLLM} %\footnote{https://github.com/vllm-project/vllm}
engine
\cite{kwon_efficient_2023}. 
For any incoming completion prompt, this library waits about 10ms for to see if it can be batched together with other requests. 
If our filter takes less than 10ms, it means we incur 0 additional latency over the base wait time. And, even if filtering takes longer, it is still possible to terminate the token-by-token generation process early, saving considerable compute as our completions take about 300--400ms to generate.
% This study thus focuses on those rejected completions, which could be helpful to the developer.
% copilot does not support manual completions as far as we know; 
% and, I have several times actually wanted a completion. 

% Moreover, the pattern of completions seems to not align well with users; 
% while shorter completions are often present, it generates extremely long ones when I have been idle for a while. 

\subsection{Joint Optimisation Objective}
\label{sec:problem:joint-optimisation-objective}
Our goal is to train a filter to invoke the completion model only at the instant a developer wants to see a completion. 
% By aligning for this, developers prefer to use our tool and we keep them in their flow. 
We partially mitigate the issues arising from optimising for acceptance rate,
%~\cite{}, 
through the observation that we may actually want to display the rejected completions that were manually invoked via a key-bind. 
By training a filter with this objective, 
along with any completions that were accepted, 
we hope to better align the completions that pass through the filter with what end-users want at that moment. 
Our objective is thus perpendicular to existing work in this area~\cite{sun_dont_2023,mozannar_when_2023}. 

The objective jointly optimises the following: 
(1) we aim to minimise the amount of times an end-user has to manually invoke the model, and
(2) we aim to minimise the amount of rejected completions that were automatically triggered. 
In other words, we consider all manual invocations and automatic, accepted completions to be our positive class (not filtered out), and rejected automatic completions to be our negative class (should be filtered). 

An added bonus of considering manually-invoked, yet rejected completions in the positive class is that the resulting filter will be less dependent on the completion-model's capabilities. 
We assume that a manual trigger is a strong indicator that a user would like to see a suggestion, and choose not to depend on whether the suggestion was accepted in this scenario.  
% as it indicates where a user would like to see a suggestion, rather than depending on the generated completion's acceptance. 
% This is relevant because our completion models are out of date compared to the state-of-the-art. 
% We will soon update them, but still want to leverage the usage data from developers up to this point to improve usability. 

\section{Approach}
\label{sec:approach}
We aim to filter out suggestions by teaching a model to predict, at any cursor location, whether to invoke the completion model or not. 
To do this, we propose to leverage our collected dataset of code completions and accompanying in-IDE telemetry to better discern the nature of interactions. 
Noting the state-of-the-art contextual understanding that transformer models exhibit, 
we explore architectures for integrating telemetry features with code context.

\subsection{{\jonberta} Architecture}
\label{sec:jonberta-architecture}

We augment a code-pretrained  
\textsc{RoBERTa}
architecture
\cite{liu_roberta_2019}, 
yielding the following two models of 84M parameters. 
As there are many possible extensions to a transformer model, we limit our search space to parameter-efficient implementations. 
Specifically, both our models incur less than 1M additional parameters. 
We assume readers' familiarity with the transformer architecture~\cite{vaswani_attention_2023}. 

\begin{itemize}
  \item {\jonbertahead} incorporating telemetry features directly in the classification head. 
  \item {\jonbertaattn} attending to (small) learned feature embeddings in the self-attention modules. 
\end{itemize}

We use the \textsc{Jon} prefix to refer to \textit{J}ointly \textit{o}ptimised atte\textit{n}tion, to both code context and telemetry data.
The motivation behind this approach lies in the state-of-the-art \textit{contextual} understanding transformer models exhibit, which we hypothesise can also leverage contextual telemetry data. 
% TODO: 'Contextual' sota understanding may need refernce, or make it clearer it refers to TEXTUAL/LINGUISTIC understanding. 
We further propose a novel tokenisation strategy centred on the cursor, to capture the most significant parts of code context.  

\begin{figure}[]
  \centering
  \includegraphics[width=0.99\linewidth]{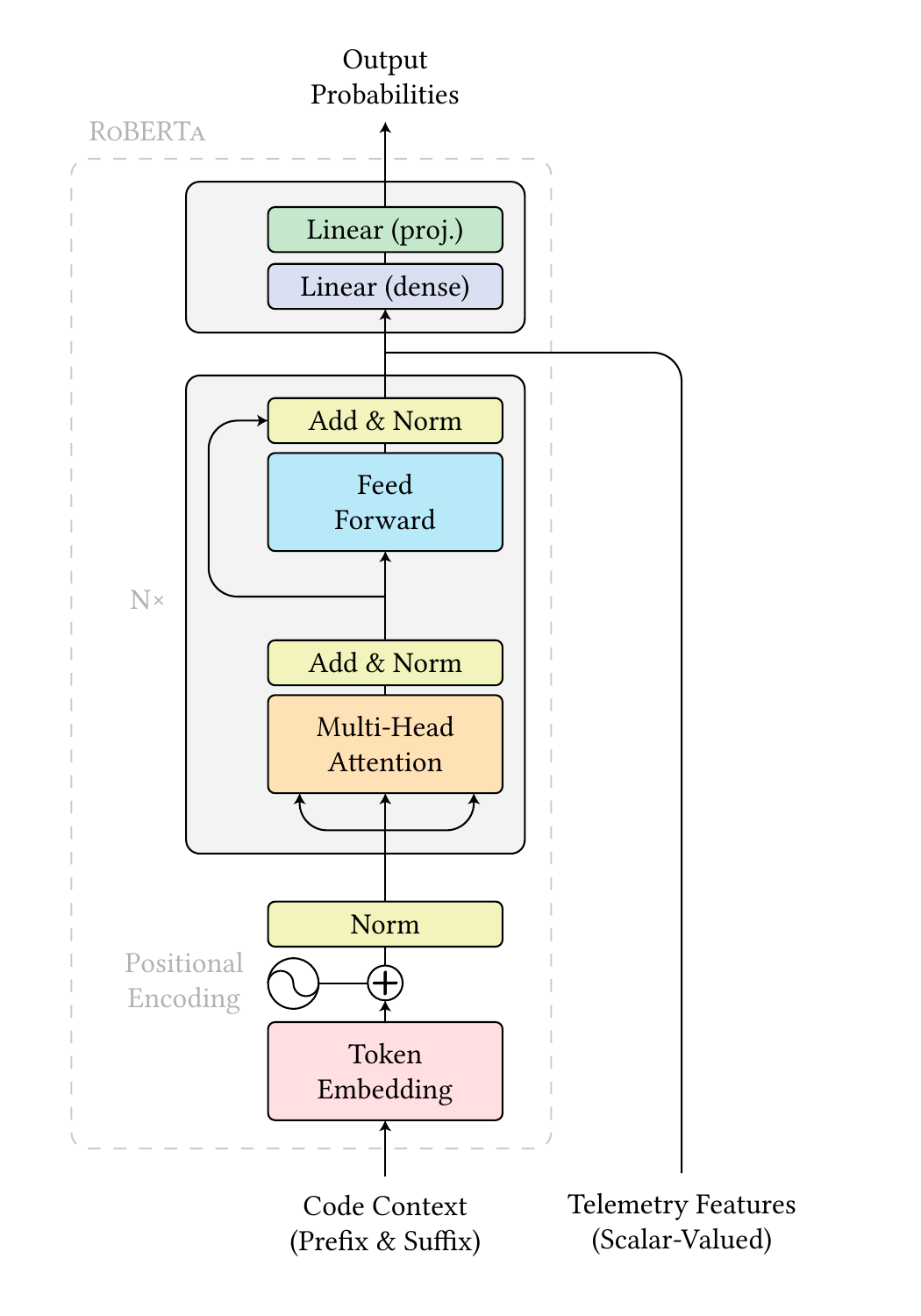}
  \caption{Telemetry Feature Data in Classification Head.}
  \label{fig:extended-classification-head}
  \Description[
    Scalar telemetry features are concatenated with the first token (special classification token) out of the last transformer block. 
  ]{
    Concatenating scalar telemetry features gives a vector. 
    The output from the last transformer block is a matrix of transformed token embeddings. 
    The classification head uses the first token's embedding to determine the class.
    We append the telemetry vector to the first token's embedding, directly before it reaches the classification head. 
  }
\end{figure}

\subsubsection{Extended Classification Head}
\label{sec:extended-classification-head}
We explore a simple {\jonbertahead} model depicted in \Cref{fig:extended-classification-head}. 
Given that a classification head first pools the output of the previous layer to the first token (\texttt{<cls>}, the classification token), it is trivial to extend this token's one-dimensional embedding with additional feature data. 
% In essence, we train a logistic classifier on the token embedding concatenated with feature data. 

In the scope of this paper, we only consider a {\jonbertahead} where the embedding is concatenated before reaching the dense layer. 
The dense layer is a matrix of size $c \times c$, where $c$ is the length of a token embedding. 
Dense layers can help the model learn low-rank embeddings~\cite{huh_low-rank_2023} 
of both feature and code context
(which can help with train/test generalisability), 
while the projection layer afterward serves as a logistic classifier. 
We increase only the dense layer's size, along one axis, 
to accommodate the concatenated features, and reinitialise it.  

\begin{figure}[]
  \centering
  \includegraphics[width=0.97\linewidth]{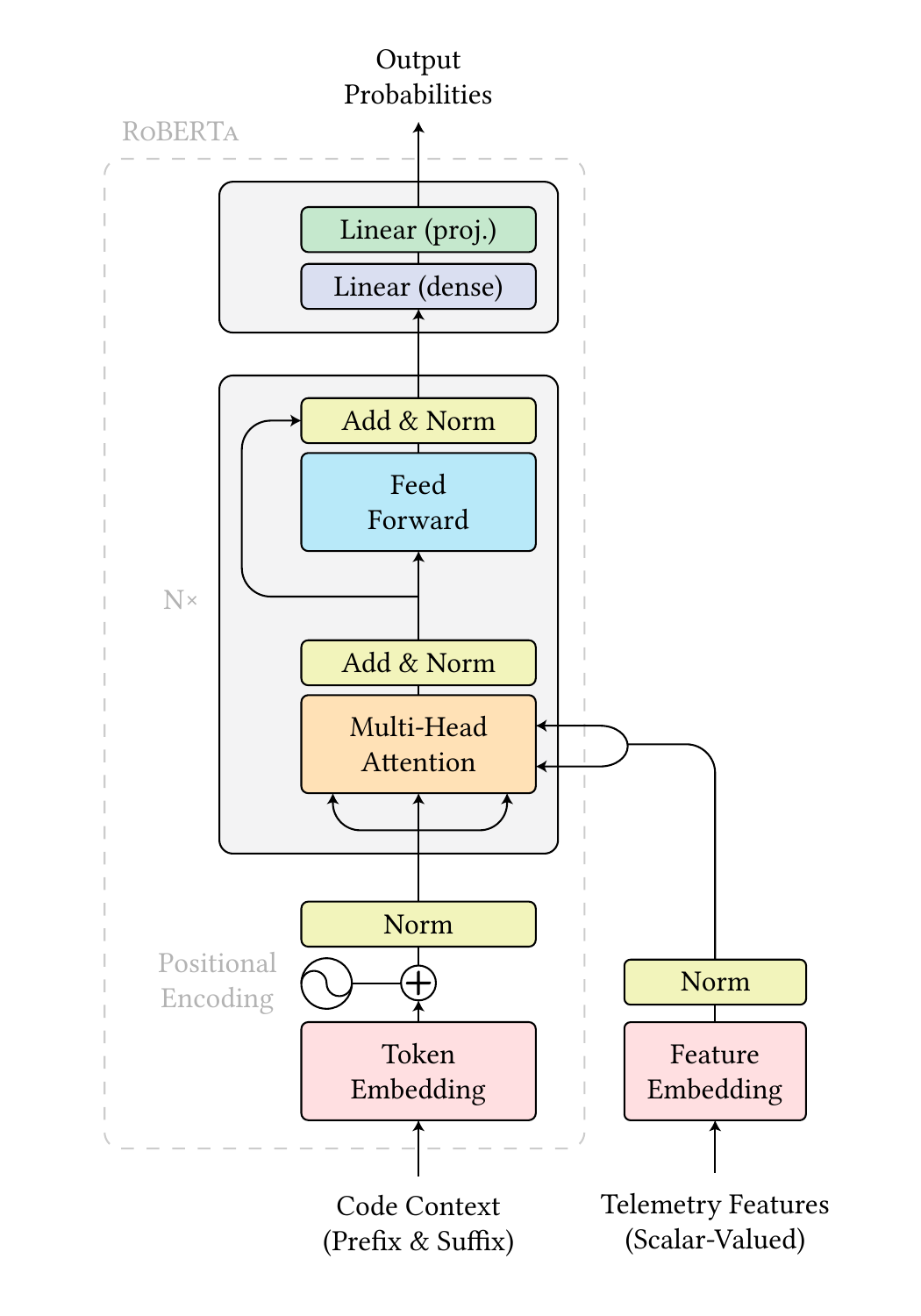}
  \caption{Self-Attention Extended to Telemetry Feature Data}
  \Description[
    We extend the attention module to also pay attention to learned feature embeddings, which are defined as functions of the scalar telemetry features. 
  ]{
    To convert any scalar feature to an embedding vector, we learn a linear function for each entry in the embedding vector. 
    We then normalise our feature embeddings using LayerNorm. 
    This allows us to generate keys and values from the feature embeddings, which can communicate with the queries from the typical tokens. 
  }
  \label{fig:extended-self-attention}
\end{figure}

\subsubsection{Extended Self-Attention}
\label{sec:extended-self-attention}

\Cref{fig:extended-self-attention} depicts our {\jonbertaattn} model, which learns feature embeddings to be attended to in the pre-existing self-attention module. 
Each weight in the feature embedding matrix is learned as a function of the corresponding scalar feature. 
Given this embedding, keys, and values can be produced to be attended to by code tokens. 
In practice, the attention module itself is equivalent to the original model, except that the keys emitted from features can dot-multiply with the queries from token embeddings, to produce weighted scores for how much a given feature's value should be added to that token embedding. 

By including telemetry feature embeddings in its self-attention mechanism, 
we hypothesise the model is able to combine both modalities 
to grasp a firmer picture of the current user intent. 
We further explore a variety of layer combinations 
and feature embedding dimensions in the online appendix of our replication package, %~\footnote{\reppackage{}}, 
but cannot conclusively state which achieves better results. 
We use a feature embedding dimension of $204$ throughout this study 
to limit the additional parameters to the model, 
while ensuring enough expressivity.

\subsubsection{Tokenisation Strategy}
\label{sec:tokenisation-strategy}

The code context provided to our {\jonbertahead} and {\jonbertaattn} models consists of the prefix (before the cursor), and the suffix (after the cursor).
Commonly, tokenisers truncate such sequence pairs either both on the left or both on the right in paired sequence-classification tasks (e.g., question-answer matching). 
However, we hypothesise that it is optimal to centre the context window on the cursor location. 
Perhaps surprisingly, something we have not yet seen in previous work.  

To achieve this within the {\jonberta} context window of 512 tokens, 
we first tokenise the suffix with right-truncation up to a maximum of 128 tokens. 
Denote the number of tokens in the suffix by $n_s$, 
which may be less than 128 if the developer is close to the end of the file. 
We then tokenise the prefix, up to a maximum of $512 - 1 - n_s$; subtracting one off the total context window here, 
to allow us to insert a \texttt{<sep>} separator token at the cursor position. 
If the total length happens to be shorter than the context window, 
we conventionally right-pad the remainder of the sequence. Note that this doesn't perfectly centre the cursor in the context window, and assumes the prefix holds more weight than the suffix. 
This decision is substantiated by the results in our online appendix. 

% While common tokenisers' truncation strategy is to either r- or l-strip 
% \ref{citation-needed, see existing work}, 
% we hypothesise that it would be optimal to centre the context window on the cursor, 
% given that the base model \textsc{RoBERTa} is an encoder. 
% We empirically motivate this in 
% \Cref{sec:tokenisation-comparison},
% where we compare prefix (before the cursor) and suffix (after) tokenisation strategies. 

\subsection{Dataset}

We train our models on data collected from a code completion plugin developed at our institution (\Cref{sec:problem:code-completion-data-and-constraints}). 
Any given code suggestion is either manually invoked via a key-bind, or automatically on a predefined set of common trigger characters (e.g., a full stop, or opening paranthesis)
\cite{izadi_language_2024}. 
As stated in our objective \Cref{sec:problem:joint-optimisation-objective}, we aim to optimise for those completions that are either (1) manually invoked, or (2) automatically invoked and accepted. 

While we have over 1M invocations of our tool, after filtering
for high-quality samples containing code context (collected on an opt-in basis, $\sim$200k samples),
and balancing our dataset by under-sampling, we maintain 
only about 10k code suggestions for training. 
Our test set, following the real-world distribution of manual and automatic invocations, contains about 20k samples,
% (but only 10\% of the smallest class used in training), 
completely separate from the training set. 
We empirically compare different dataset distributions in our online appendix, motivating the under-sampled training distribution in this classification scenario where classes are not equally represented in practice. 

\begin{table}[h]

  \caption{Class Distribution in Our Code Completion Train/Validation and Test Datasets.} 
  \label{table:dataset-distribution}
  
  \begin{tabular}{lrr|r}

    \toprule

    Class             & Positive &                    & Negative        \\
    \midrule 

    Completion        & \m{Manual}     & \mc{Automatic}                   \\ 
    Type              &                & Accepted        & Rejected       \\
    \midrule

    Test              & 6118 (27.8\%)  & 431   (1.9\%)   & 15889 (70.3\%) \\
    Train/Validation  & 3909 (33.3\%)  & 3909 (33.3\%)   & 3909  (33.3\%) \\

    \bottomrule
  \end{tabular}

\end{table}

\Cref{table:dataset-distribution} shows the (sub-)class distribution of our train/validation and test dataset. 
We purposefully avoid distinguishing between manual accepted, and manual rejected invocations of our tool, as we consider both to be part of our positive class. 
As added context, however, the manual rejected invocations constitute a total of 47\% of our positive class in the real-world (test) distribution. Unfortunately, we are unable to share our collected dataset as it contains user-sensitive code context.

\section{Experimental Setup}
\label{sec:experimental-setup}

We laid out the code completion plugin context and problem constraints in \Cref{sec:problem-definition}. 
And, having established our search space for this problem in \Cref{sec:approach},
we now propose how to navigate it. 

\subsection{Research Questions}
\label{sec:research-questions}

As defined in \Cref{sec:problem:joint-optimisation-objective}, 
the objective of our models is to label 
manual and automatic invocations which are accepted as positive (helpful); 
and, consider the remaining automatic invocations that are rejected 
as the negative class (unhelpful). 
To this end, we pose the following research questions: 

% To filter out unhelpful completions, we train models on telemetry data and code context, and explore hybrid architectures. We evaluate the resulting models on their prediction accuracy on the classes identified in \Cref{sec:problem:joint-optimisation-objective}. Lastly, we investigate differences between offline and online evaluation.

\begin{itemize}

  \item[RQ1] \textbf{How does a transformer model compare to a baseline logistic regression model at filtering out unhelpful suggestions (offline evaluation)?} 
  We fine-tune a code-pretrained \textsc{RoBERTa} 
 ~\cite{liu_roberta_2019} model (\textsc{CodeBERTa}) on code completion snippets collected from our plugin, and evaluate it against a logistic regression baseline trained on telemetry and selected textual features. The baseline is inspired by reverse-engineering GitHub Copilot and reflects the state-of-the-art.

  \item[RQ2] \textbf{Can a pre-trained transformer be extended to incorporate an additional modality consisting of telemetry feature data?} 
  We train our {\jonberta} variants (\Cref{sec:jonberta-architecture}) with telemetry feature data as an additional modality when making predictions. 
  % We augment the fine-tuned \textsc{RoBERTa} filters to also include telemetry feature data when making predictions. We define a search space, and compare two approaches (\Cref{sec:jonberta-architecture}) to incorporate this extra modality in its architecture. 

  \item[RQ3] \textbf{How effective are the above approaches in a real-world setting (online evaluation)?} 
  We deploy our filters in a code-completion plugin to investigate whether the filters' decisions align with \usercount users in practice. We further evaluate the computational feasibility of our approach, and note discrepancies between the offline and online environments. 

\end{itemize}

\subsection{Evaluation Settings and Metrics}
\label{sec:evaluation-settings}

\subsubsection{Metrics for Offline Settings (RQ1 \& RQ2)}
\label{sec:evaluation-offline-metrics}

To evaluate whether our models can capture the different invocation types that determine our classes (see \Cref{sec:problem:joint-optimisation-objective}), we compute accuracy per manual, accepted automatic, and rejected automatic subclass. Based on this, we further compute the macro average accuracy, to serve as a single metric to compare models on.

We choose macro average accuracy (across classes), as opposed to micro average (across all samples), as our positive classes are under-represented in our code-completion dataset. 
As a result, it is paramount that completions the developer wants to see are prioritised against the vast majority of completions that are ignored. 
We assume that the mistake of filtering out a completion when a developer would want to see one is worse than showing a completion when the developer does not want to see one. 

As shown in \Cref{table:dataset-distribution}, some of our classes are considerably under-represented. The variance due to such a small dataset can become pronounced when training transformer models. To capture this variance, we train five models on five train/eval splits (9:1). Then, we bootstrap our accuracy scores on the test set by alternatingly taking a sample from each of the five models, 
for a total of $n = 10,000$ samples.

\subsubsection{Metrics for Online Setting }
\label{sec:evaluation-online-metrics}

For our online evaluation, we no longer have a valuable distribution of manual/automatic classes as we remove the predefined trigger-point invocation rule. 
To remedy this, we evaluate completions that pass the filter via (1) acceptance rate as a proxy for their \textit{timing} with developers' mode of thought; 
and (2) score accepted completions using CodeBERTScore
~\cite{zhou_codebertscore_2023} 
as a proxy for their \textit{quality}. 
We also measure the latency in milliseconds. 

CodeBERTScore is a recently-proposed measure that correlates closest with both functional correctness and human preference
~\cite{zhou_codebertscore_2023}.
This is contrary to the oft-seen CodeBLEU, METEOR, and ROUGE-L measures which are designed for natural languages and do not work well with the syntactic structure of programming languages
~\cite{evtikhiev_out_2023}. 
CodeBERTScore is computed by passing the code completion and ground truth (after 30s) through a code-pre-trained encoder model,
and then computes the F3 score based on the similarity between token embeddings at a layer that correlates best with human preference and functional correctness.

We further propose the harmonic mean of these two as a single metric to compare models by.
Specifically, optimising only acceptance rate results in worse completions
\cite{ziegler_productivity_2022,mozannar_when_2023}. 
And, optimising just the content of a completion, does not make the filter align well with developers, as evidenced by the number of manual completions we observe in our dataset.  
We hope this communicates to the reader how these two measures should be weighed in our framing of the problem. 

\subsection{Feature Engineering and Baselines}
\label{sec:feature-engineering}

The features extracted from our code-completion data for the Logistic Regression, \textsc{CodeBERTa}, and {\jonberta} models, are shown in \Cref{table:extracted-features}. 
We purposely avoid providing {\jonberta} with features that can be inferred from the code context (e.g., whether there is whitespace after the cursor), to assert it is able to leverage that data implicitly. 
To this end, we define three types of feature data: 

\begin{itemize}
  \item[T] Telemetry as those features that cannot be directly extracted from a snippet of code. 
  % Unintuitively, this means things such as the document length itself, as it may be longer than the snippet which is constrained by the model's context window. \\
  \item[C] Code context as those \textit{textual} features that are explicitly extracted by fixed rules. 
  \item[S] Snippet as the prompt to the completion model, truncated to the filter model's context window (512 tokens) 
  using our centred-on-cursor strategy (\Cref{sec:tokenisation-strategy}). 

\end{itemize}

\begin{table}[h]

  \caption{Features Used in Classification, per Filter Model.}
  \label{table:extracted-features}
  \begin{tabular}{ll|rrr}
    \toprule 
    &                      & Log. Reg.               &\textsc{Code} &\textsc{Jon} \\
    \midrule                                                             
    % \\
    % data                                                               
    % Previous filter label       & \checkmark                   \\        & none      
    T1 & Time since last completion  & \checkmark              &              & \checkmark       \\
    T2 & Document length             & \checkmark              &              & \checkmark       \\
    T3 & Cursor offset               & \checkmark              &              & \checkmark       \\
    T4 & Offset as percentage        & \checkmark              &              & \checkmark       \\
    T5  & language (20 options)      & \checkmark              &              & \checkmark      \\
    T6 & IDE ( \texttt{jetbrains} / \texttt{vscode} ) 
                                     & \checkmark              &              & \checkmark      \\
\midrule
    C1 & Length of last prefix line  & \checkmark              &              &                  \\
    C2 & Above, without whitespace   & \checkmark              &              &                  \\
    C3 & Whitespace after cursor     & \checkmark              &              &                  \\
% \midrule
    C4 & Last prefix char (ASCII 32-125) & \checkmark          &              &                 \\
    C5 & Above, without whitespace   & \checkmark              &              &                 \\
\midrule
    S1 & Prefix                      &                         &   \checkmark & \checkmark      \\
    S2 & Suffix                      &                         &   \checkmark & \checkmark      \\
    \bottomrule

  \end{tabular}
\end{table}

The features for the logistic regression model are inspired by reverse-engineering Copilot~\cite{thakkar_copilot_2023}, with the exception of one feature that depends on a pre-existing filter (which we do not have). 
The languages are the same as the 20 considered by Copilot.
We choose to follow Copilot as, to our knowledge, this is the only filter currently deployed in practice, and contains most of the significant features found in previous work~\cite{mozannar_when_2023}. An extended explanation of these features can be found in our online appendix. 

% The 20 language options are adopted from previous work~\cite{izadi_language_2024,xu_systematic_2022}. But, as they do not perfectly align with the most common languages our \pluginname plugin is used with~\cite{izadi_language_2024}, this leaves room for further fitting to our data distribution. 

\subsection{Configuration and Implementation Details}
\label{sec:config-and-implementation-details}

We fine-tune all transformer models for six epochs 
with a $2e^{-5}$ learning rate and $16$ batch size from a public 
\textsc{CodeBERTa-base-v1}\footnote{\label{foot:codeberta-small-v1}https://huggingface.co/huggingface/CodeBERTa-small-v1}
checkpoint, with each epoch containing about 10k training samples. 
We also use its tokeniser for snippet features. 
For the {\jonberta} models, 
we fine-tune from the 3rd-epoch \textsc{CodeBERTa} checkpoint, 
for an additional three epochs; 
as we observe this results in stabler training than training from the public checkpoint.
% We publish our fine-tuned \textsc{CodeBERTa} and {\jonberta} models on HuggingFace~\footnote{https://doi.org/10.5281/zenodo.10935342}, 
% and a replication package~\footnote{\reppackage{}}. 

All of our transformer models are implemented with 
\texttt{PyTorch}\footnote{https://pytorch.org}, 
and trained on an NVIDIA GeForce RTX 3080 GPU, 
taking about $20$ minutes per model. 
All metrics are computed using the functions provided by 
\texttt{scikit-learn}~\footnote{http://scikit-learn.org} library. 
For our online evaluation, 
we use an inference server with an NVIDIA GeForce RTX 2080 Ti, 
separate from our training setup. As this is a relatively older GPU, 
we expect slightly higher latency during filter inference.

\section{Results}
\label{sec:results}

% \subsection{RQ1: Textual Information Drives Performance}
\subsection{RQ1: Impact of Code Context}
\label{sec:results-rq1}

To address RQ1, we evaluate the contributions from training on 
snippet features (S), 
against telemetry (T), 
and fixed-rule textual features (C).
To this end, we first consider logistic regression baselines 
trained on iteratively more T and C (see \Cref{sec:feature-engineering}). 
% The baselines are inspired by Copilot, as to our knowledge it is the only filter currently deployed in practice. 
Lastly, \pluginname is cross-application, so we include a feature for the IDE. 
In our dataset, JetBrains users tend to have a higher suggestion acceptance rate, partially because they support only the popular languages~\footnote{Very recently, JetBrains did release a preview for a cross-language IDE, Fleet: https://www.jetbrains.com/fleet/} which code completion models tend to perform best on 
\cite{katzy_impact_2023}. 

We do not directly extract the weights from its plugin code.
Instead, we retrain the Copilot-style baselines on our own completion-request dataset.
because the data distribution of our completion plugin is likely different
due to its different invocation methods and completion style as explained in \cref{sec:problem:code-completion-data-and-constraints}.

% (this means that all of our features are likely dependent on this one to some extent, which breaks the assumptions of a logistic regression classifier; but copilot also does not guarantee that their features are independent.)

\begin{table}[h]

  % TODO: just give the median values instead of the mean;
  % which should shrink the error bounds slightly

  \caption{Filter accuracy for Logistic Regression and \textsc{CodeBERTa} classification models, given per invocation sub-class.} 
  \label{table:filter-accuracy-logres-codeberta}

  \begin{tabular}{lrrrr}
    \toprule
                            & Manual          & Auto/acc.       & Auto/rej.       & \textbf{Avg.}     \\
                            % &$n=6118$         &$n=431$          &$n=15889$        &                    \\
    \midrule                                                                    
                                                                                
    Logistic Regr.                                                                                     \\
    \quad Telemetry $_{\text T1-5}$ & 99.6 $\pm$0.3  & 99.1 $\pm$0.7  &  1.4 $\pm$0.7  & 66.7          \\
    % TODO: could 
    \quad + Textual $_{\text C1-4}$ & 98.6 $\pm$0.3  & 66.8 $\pm$4.5  & 61.9 $\pm$1.2  & 75.8          \\
    \quad + Copilot $_{\text C5 }$  & 98.6 $\pm$0.3  & 66.1 $\pm$4.6  & 63.9 $\pm$0.9  & 76.2          \\
    \quad + IDE     $_{\text T6 }$  & 98.5 $\pm$0.3  & 66.1 $\pm$4.6  & 65.0 $\pm$0.9  & 76.5          \\

    % 3 epochs
    % \textsc{CodeBERTa}      & 98.3 $\pm$ 0.5  & 75.7 $\pm$ 7.6  & 68.2 $\pm$ 3.0  & 80.7 \\
    % 6 total epochs (like RQ2)
    \textsc{CodeBERTa}      & 98.5 $\pm$0.4  & 74.7 $\pm$4.6  & 73.1 $\pm$1.2  & \textbf{82.1}   \\

    \bottomrule
  \end{tabular}

\end{table}

As shown in \Cref{table:filter-accuracy-logres-codeberta}, 
a model trained on telemetry features (T1--5) alone, while attaining average accuracy of 66.7\%, is completely unable to distinguish automatic invocations that end up being rejected (1.4\% accuracy). 
Thus, it is necessary to include some explicit textual features (C1--4) for distinguishing these classes.
This intimates that extended code context can be leveraged. 
We refer to our replication package for additional experiments with different granularities of textual features. 

Notably, \textsc{CodeBERTa}, trained on solely code snippet (S) features, can outperform the best baseline on automatically accepted and automatically rejected queries by $9.6$ and $3.2$ absolute percentage points, respectively.
This indicates that the semantic understanding of code such a transformer model exhibits 
propels it past the classification baseline. 
And, furthermore, this snippet modality likely contributes orthogonally to the telemetry data, as both are distinct features that cannot be inferred from each other. 
This motivates our architectural exploration to attend to both these modalities in one classification model.

% \subsection{RQ2: Hybrid {\jonberta} Models Are Promising}
\subsection{RQ2: Hybrid {\jonberta} Models}
\label{sec:results-rq2}

To address RQ2, we train {\jonberta} models leveraging both snippet and telemetry modalities 
and compare them to our new \textsc{CodeBERTa} baseline. 
This architectural search space is especially vast for the {\jonbertaattn} model, 
due to the inclusion of feature \textit{embeddings} with tuneable parameters.
As such, we defer most of our experiments to the online appendix in our replication package, as well as a few {\jonbertahead} experiments. 
We choose to only display the first-layer configuration here\footnote{{\jonbertaattn} 0L: Ranking 19/67, Median average score is 81.7.}
to give telemetry embedding an equally early chance as token embeddings at communicating in the attention mechanism;
and choose an embedding dimension of $204$ to not incur too many additional parameters. 

\begin{table}[h]

  \caption{Filter accuracy for \textsc{CodeBERTa} and {\jonberta} classification models, given per invocation sub-class. 
  Error Bounds are for $p < 0.5$ via Bootstrapping $n = 10\;000$.} 
  \label{table:filter-accuracy-code-jonberta}
  
  \begin{tabular}{llrrrr}
    \toprule
                                  && Manual        & Auto/acc.     & Auto/rej.     & \textbf{Avg.}    \\
                                  % && $n=6118$      &$n=431$        &$n=15889$      &                \\
    \midrule                                                                    

    % \textsc{CodeBERTa} (3 epoch) & 98.3 $\pm$ 0.5 & 75.7 $\pm$ 7.6 & 68.2 $\pm$ 3.0 & 76.5 $\pm$ 2.1 & 80.7 \\

    % 6 epoch
    \textsc{CodeBERTa}            && 98.5 $\pm$0.4           & 74.7 $\pm$4.6          & \textbf{73.1} $\pm$1.2 & 82.1 \\

    {\jonberta} \\                                                                
    \quad \textsc{head} (dense)   && \textbf{98.6} $\pm$0.4  & \textbf{78.0} $\pm$6.2 & 71.4 $\pm$5.1          & \textbf{82.7} \\
    \quad \textsc{attn} (0L)      && \textbf{98.6} $\pm$0.4  & 75.0 $\pm$6.8          & 72.5 $\pm$3.5          & 82.0 \\
    
    \bottomrule
  \end{tabular}

\end{table}

% While our results in \Cref{table:filter-accuracy-code-jonberta} are less conclusive than before, 
In \Cref{table:filter-accuracy-code-jonberta}, {\jonbertahead} promisingly shows it can better discern between completion classes, though our results are less conclusive than before. 
While incorporating the telemetry features in the first attention layer of {\jonbertaattn} matches the performance of {\jonbertahead}, 
implying it can learn to integrate telemetry features with its semantic understanding of code, this may equally well be attributable to variance.
This could be, in part, due to the limited training set size as revealed by our bootstrap strategy for computing error bounds (\Cref{sec:evaluation-settings}), 
which we further discuss as an internal threat to validity in \Cref{sec:internal-validity}.

Regardless, we present these results motivated by the same argument as in \Cref{sec:problem-definition}:
Transformer models are becoming increasingly integrated into software engineering tasks, but also well outside of the field.
As a result, there is a need to integrate additional modalities into pre-trained models, 
as tokens are but one of many sources of information that can be leveraged in AI-powered tools.

% \subsection{RQ3: Online Evaluation Confirms Applicability}
\subsection{RQ3: Online Evaluation}
\label{sec:results-rq3}

To evaluate how our filters fare in the real world, we deploy them in a code completion plugin with \usercount developers over a \timeperiod period, resulting in \completioncount requests. 
To this end, we disable the predefined trigger-point constraint that was described in \Cref{sec:problem:code-completion-data-and-constraints}, to now automatically invoke the completion model at the historically manual trigger-points.

We perform an A/B study by assigning each user one of the following five filters per coding session. 
We define a session as a sequence of completion requests where any two are no more than 30 minutes apart, to avoid end-user confusion from different completion behaviour on every request. 
For all filters, requests with a prompt (prefix + suffix around the cursor) less than 10 characters are automatically rejected, as they do not have enough context for a worthwhile completion. 

\begin{itemize}
  \item[1] \textbf{None}: all completion requests pass through. 
  \item[2] \textbf{Logistic regression} using telemetry (T) and context (C) features. 
  \item[3] \textbf{\textsc{CodeBERTa}} using only snippet (S) features.
  \item[4 \& 5] \textbf{{\jonbertahead}} and \textbf{\textsc{-attn}} using both T and S. 
\end{itemize}

\begin{table}[h]

  \caption{Completion statistics for filters deployed in \pluginname. The relative acceptance rate is with respect to None. Latency is computed as the median of all requests.} 
  \label{table:online-filter-stats}
  
  \begin{tabular}{lrrrrr}
    \toprule

    \m{Filters}        &            &           &                   & \mc{\jonberta}                 \\
                       & None       & LogReg. & \textsc{Code}     & \textsc{head} & \textsc{attn}  \\
    \midrule

    \textbf{Requests} \\
    \; Received        & 13.5k      & 10k       & 20k               & 15k           & 13k             \\
    \; Filtered out    &  2.2\%     & 29.4\%    & 36.6\%            & 34.4\%        & 39.4\%          \\

    \textbf{Completions} \\
    \; Shown           & 97.8\%     & 70.5\%    & 63.2\%            & 65.6\%        & 60.5\%          \\
    \; Accepted        & 1.42\%     & 0.62\%    & 1.20\%            & 1.44\%        & 1.03\%          \\

    \textbf{Accepted}  \\
    \; Relative rate   & 100.0\%    & 45.0\%     & 84.5\%           & 101.0\%          & 72.3\%       \\
    \; CodeBERTScore   & 0.76       & 0.94       & 0.85             & 0.82             & 0.88         \\
    \; Harmonic Mean   & 0.864      & 0.609      & 0.847            & \textbf{0.905}   & 0.794       \\

    Latency (ms)       & 0.0        & 0.2       & 25.0              & 20.8          & 24.1            \\

    \bottomrule
  \end{tabular}

\end{table}

\Cref{table:online-filter-stats} shows our results.
Using our proposed harmonic mean to convey the balance of suggestion quality and timing, {\jonbertahead} performs best. 
While our proposed metric maintains ordinality among the \textsc{CodeBERTa} and {\jonberta} models, compared to the offline evaluation, this also highlights our proposed metric is not perfect, as presumably, all our filters should be performing better than the no-filter baseline. 
We defer discussion on alternate weightings for the harmonic mean to future work 
(\Cref{sec:construct-validity}), 
to avoid conflicting interests from tuning our metric here. 

% As we remove the predefined trigger-points from our plugin, we change its data distribution and thus cannot fairly compare it to the historical data. 
% Furthermore, while we do still allow users to manually invoke a completion (bypassing the filter), we do not have enough samples of this to draw any conclusive results ($<30$ out of the \completioncount total), 
% other than that our filters do not prevent completions where the user wanted one. 

Additionally, we demonstrate the feasibility of our approach in practice. 
A lightweight transformer model can be deployed server-side as a filter for incoming requests, with relatively minimal latency compared to the completion model itself, which takes 300-400ms in our case. 
Future work can consider further optimising these models, through e.g., model compression 
\cite{wu_extreme_2022}, for client-side deployment. 

%% NOTE: The Acceptance Rate statistic in the \pluginname paper is computed after the following filtering steps:
%  1. Exclude predictions where the ground-truth is empty (after 30s)
%     (this is unfair imho, as it can easily be the case that the developer went looking for docs)
%  2. Exclude invocations where the predictions are empty 
%     (models are incapable / preds are filtered following a bunch of rules specified all over the codebase)
%  Both of these fall under the 'unnecessary invocations' umbrella, 
%  which are the ones I want to filter out with my models. 

\section{Discussion}
\label{sec:discussion}

\mali{We anticipate that the issue of redundant invocation will become even more noticeable as software engineering tasks increasingly incorporate billion-parameter transformer models.}
Our results demonstrate the effectiveness of using a smaller, 
lightweight transformer to control when a larger, completion model is invoked. 
% \aral{Furthermore, we believe this problem is more general than transformer-backed code completion}
\mali{Moreover, we believe smart invocation filtering models such as ours 
not only enhance code completion but also any other transformer-based interaction with users.}

To our knowledge, we are the first work to augment a \textit{pre-}trained transformer with additional feature modalities.
Transformer models exhibit exceptional contextual understanding, yet are bottlenecked by the textual medium.
Especially considering that in-app telemetry data is often collected anyway, 
we highlight that it is fruitful to leverage this additional input dimension.
By showing promising results in this search space, 
we hope to inspire others to venture deeper. 

\balance

\subsection{Threats to Validity}
\label{sec:threats-to-validity}
\subsubsection{Internal: Limited Contextual Usage Data}
\label{sec:internal-validity}
% \textbf{Internal Threats}
% \label{internal-validity}

% Our training dataset consisting of only 10k samples is not ideal for exploring the search space of hybrid transformer models fine-tuned on additional feature data. 
% It is likely a fruitful direction for future work to thoroughly navigate this search space using a larger dataset, to be able to draw more significant comparative results between architectures.
\mali{In RQ1 and RQ2, we used a training dataset of just 10k samples. This size is not optimal for fully examining the potential of hybrid transformer models enhanced with extra feature data. Expanding the dataset in future research would likely offer a better understanding of these models' capabilities, enabling more meaningful comparisons between different architectures.}

\subsubsection{External: Generalisability to Other Code-Completion Tools}
\label{sec:external-validity-other-tools}
% \subsubsection{External Threats}: 
% \label{}
% As we use a code-suggestion plugin with a relatively small user-base, 
% compared to solutions like GitHub Copilot and JetBrains AI, 
% there are several factors that may affect the generalisability of our results to other code completion plugins. 
% We already establish differences in \Cref{sec:problem:code-completion-data-and-constraints}.
\mali{We utilized a code-suggestion plugin that has a smaller user base compared to larger production systems. This choice introduces several factors that might impact how our findings can be applied to other code completion tools. These differences have been detailed in \Cref{sec:problem:code-completion-data-and-constraints}. Our approach, while specific, offers valuable insights but warrants caution when generalizing to other contexts.}

%%
% Our completion plugin is designed for line-completion, 
% while other solutions tend to generate any number of tokens. 
% This additional constraint theoretically makes our problem simpler by definition, 
% however, also may lead to less-generalisable results. 
% Furthermore, the UI of our plugin differs from the conventional ghost-text display,
% in that suggestions are shown alongside language-server recommendations in a completion box. 
% It may be that the nature of developer interactions are different in this context, 
% again leading to a loss of generalisability. 
%%
% Furthermore, our plugin combines suggestions from three backend models: 
% Incoder, UniXcoder, and CodeGPT, displayed in random order to the end-user. 
% Given that these models have different capabilities, and that 
% Incoder achieves roughly double the acceptance rate of the other two 
%~\cite{izadi_language_2024}, 
% this is a source of variance in our dataset. 

\subsubsection{Construct: Limitations of the Proxy Metrics}
\label{sec:construct-validity}
% \textbf{Construct Threats}:
% We use the harmonic mean of acceptance rate and CodeBERTScore~\cite{zhou_codebertscore_2023} 
% in our online evaluation. Future research can apply and investigate this metric and its alternative weighings. 
% And, like previous work~\cite{mozannar_when_2023,ziegler_productivity_2022}, 
% we note that our proxy metrics may be inadequate to assess this interaction usability without adversely affecting it from another angle; and further qualitative studies with developer interviews may help in this regard.
In our online evaluation, we use the harmonic mean of acceptance rate and CodeBERTScore as our metric to measure performance. This approach, suggested for further exploration in future studies, allows for adjustments in how each component is weighted. 
Consistent with earlier research \cite{mozannar_when_2023,ziegler_productivity_2022}, we acknowledge that these proxy metrics might not fully capture the usability of the interaction without potentially compromising it in some other way. To gain a deeper understanding, we suggest that future work could benefit from qualitative studies, including interviews with developers, to complement these quantitative measures.
% Showing completions when a developer wants them in that instant is not necessarily well-aligned with what developers want long term. 
% While it is difficult to track the long-term contribution of a completion, this is an issue warranting additional research as the adoption of AI tools correlates with code churn~\cite{harding_coding_2024}. 
% 

\subsubsection{Ethical Considerations}
\label{sec:ethical-considerations}

% It is worth mentioning that we do not address potential privacy-related concerns arising from training on usage data. 
% Classification models do not suffer the same issues as generative models in this area, and we consider it out-of-scope for this study. 
% Instead, we would like to point the reader towards the growing hardware support on consumer devices
% \cite{alizadeh_llm_2023}.
% And, orthogonally, the increasing research on small LMs
% \cite{warstadt_findings_2023,eldan_tinystories_2023}
% and model compression that could mitigate these concerns
% \cite{zhu_survey_2023,zhang_lifting_2023,wu_extreme_2022,de_moor_codegpt_2023}.
% \mali{For our research, we have obtained approval from our institution's ethical board, ensuring that all practices meet the required ethical standards. Additionally, we have secured explicit consent from users before collecting and using their information. In addressing potential privacy concerns associated with the use of developer usage data in our study, we have opted not to explore these issues deeply. Our decision is based on the nature of our invocation filtering models, which, unlike generative models, are less likely to raise privacy concerns by leaking data.}
\mali{Our research received approval from the institutional ethical board and explicit user consent for data use. Additionally, we have secured explicit consent from users before collecting and using their information. We chose not to deeply investigate privacy issues related to developer data usage, as our invocation filtering models (classifiers) pose fewer privacy risks compared to generative models. However, to comply with the GDPR, we are unable to share our dataset as we cannot guarantee anonymity.}

\section{Conclusion and Future Work}
\label{sec:conclusion}

\aral{To summarise, we train a transformer-based invocation-filtering model on a dataset 
we collected from an open-source code completion plugin, \pluginname. 
We show that code context is especially useful in filtering predictions, and highlight the potential of integrating this information with the telemetry data collected in an IDE.
Lastly, we deploy our filters in practice and show their practical effectiveness in both offline and online settings.}

Future work can more thoroughly explore the search space we have established, by utilising a larger dataset. 
Our limited dataset may not fully represent the diverse behaviours of developers, 
and related work shows promising results in personalising the invocation-filtering system~\cite{mozannar_when_2023,chen_gmail_2019}. 
We choose not to explore this avenue in this study to limit our architectural search space, though strongly advocate for further exploration in this area. 

Lastly, we note that delivering completions exactly when a developer requests them might not always match what developers truly need in the long run. 
Tracking the long-term impact of these completions presents challenges, yet understanding this is crucial, especially as the use of AI tools shows a link to increased code changes. 
This domain deserves further investigation to better align for lasting developer benefits.

% %%
% %% The acknowledgments section is defined using the "acks" environment
% %% (and NOT an unnumbered section). This ensures the proper
% %% identification of the section in the article metadata, and the
% %% consistent spelling of the heading.
% \begin{acks}
% I would like to express gratitude to the AISE lab members for being a source of inspiration and continual support. 
% Specifically, I would like to thank Jonathan Katzy for his suffering through my frequent questions about transformer architectures; 
% and Fabio Salerno for his contribution in statistically analysing the results.
% Of course, I am also very grateful to Maliheh Izadi for giving me this opportunity, and her students for providing the code completion dataset and implementation support.
% And lastly, in an effort to normalise psychological support to the same degree as practical and intellectual contributions, a big up to my homie Kerem for being a rock-hard pillar of heartening support.
% \end{acks}

\newpage
\nobalance
\bibliographystyle{ACM-Reference-Format}
\bibliography{main}


\begin{thebibliography}{49}

%%% the Web address.

\ifx \showCODEN    \undefined \def \showCODEN     #1{\unskip}     \fi
\ifx \showDOI      \undefined \def \showDOI       #1{#1}\fi
\ifx \showISBNx    \undefined \def \showISBNx     #1{\unskip}     \fi
\ifx \showISBNxiii \undefined \def \showISBNxiii  #1{\unskip}     \fi
\ifx \showISSN     \undefined \def \showISSN      #1{\unskip}     \fi
\ifx \showLCCN     \undefined \def \showLCCN      #1{\unskip}     \fi
\ifx \shownote     \undefined \def \shownote      #1{#1}          \fi
\ifx \showarticletitle \undefined \def \showarticletitle #1{#1}   \fi
\ifx \showURL      \undefined \def \showURL       {\relax}        \fi
% invisible to TeX
\providecommand\bibfield[2]{#2}
\providecommand\bibinfo[2]{#2}
\providecommand\natexlab[1]{#1}
\providecommand\showeprint[2][]{arXiv:#2}

\bibitem[Amazon CodeWhisperer(2023)]%
        {codewhisperer_2023}
Amazon CodeWhisperer \bibinfo{year}{2023}\natexlab{}.
\newblock \bibinfo{title}{AI Code Generator}.
\newblock \bibinfo{howpublished}{Online}.
\newblock
\urldef\tempurl%
\url{https://aws.amazon.com/codewhisperer/}
\showURL{%
\tempurl}


\bibitem[Barke et~al\mbox{.}(2022)]%
        {barke_grounded_2022}
\bibfield{author}{\bibinfo{person}{Shraddha Barke}, \bibinfo{person}{Michael~B. James}, {and} \bibinfo{person}{Nadia Polikarpova}.} \bibinfo{year}{2022}\natexlab{}.
\newblock \showarticletitle{Grounded {Copilot}: {How} {Programmers} {Interact} with {Code}-{Generating} {Models}}.
\newblock \bibinfo{journal}{\emph{Proceedings of the ACM on Programming Languages}} \bibinfo{volume}{7}, \bibinfo{number}{OOPSLA1} (\bibinfo{date}{Oct.} \bibinfo{year}{2022}), \bibinfo{pages}{85--111}.
\newblock
\showISSN{2475-1421}
\urldef\tempurl%
\url{https://doi.org/10.1145/3586030}
\showDOI{\tempurl}


\bibitem[Bavarian et~al\mbox{.}(2022)]%
        {bavarian_efficient_2022}
\bibfield{author}{\bibinfo{person}{Mohammad Bavarian}, \bibinfo{person}{Heewoo Jun}, \bibinfo{person}{Nikolas Tezak}, \bibinfo{person}{John Schulman}, \bibinfo{person}{Christine McLeavey}, \bibinfo{person}{Jerry Tworek}, {and} \bibinfo{person}{Mark Chen}.} \bibinfo{year}{2022}\natexlab{}.
\newblock \bibinfo{title}{Efficient {Training} of {Language} {Models} to {Fill} in the {Middle}}.
\newblock
\newblock
\urldef\tempurl%
\url{http://arxiv.org/abs/2207.14255}
\showURL{%
\tempurl}
\newblock
\shownote{arXiv:2207.14255 [cs]}.


\bibitem[Chen et~al\mbox{.}(2021)]%
        {chen_evaluating_2021}
\bibfield{author}{\bibinfo{person}{Mark Chen}, \bibinfo{person}{Jerry Tworek}, \bibinfo{person}{Heewoo Jun}, \bibinfo{person}{Qiming Yuan}, \bibinfo{person}{Henrique Ponde de~Oliveira Pinto}, \bibinfo{person}{Jared Kaplan}, \bibinfo{person}{Harri Edwards}, \bibinfo{person}{Yuri Burda}, \bibinfo{person}{Nicholas Joseph}, \bibinfo{person}{Greg Brockman}, \bibinfo{person}{Alex Ray}, \bibinfo{person}{Raul Puri}, \bibinfo{person}{Gretchen Krueger}, \bibinfo{person}{Michael Petrov}, \bibinfo{person}{Heidy Khlaaf}, \bibinfo{person}{Girish Sastry}, \bibinfo{person}{Pamela Mishkin}, \bibinfo{person}{Brooke Chan}, \bibinfo{person}{Scott Gray}, \bibinfo{person}{Nick Ryder}, \bibinfo{person}{Mikhail Pavlov}, \bibinfo{person}{Alethea Power}, \bibinfo{person}{Lukasz Kaiser}, \bibinfo{person}{Mohammad Bavarian}, \bibinfo{person}{Clemens Winter}, \bibinfo{person}{Philippe Tillet}, \bibinfo{person}{Felipe~Petroski Such}, \bibinfo{person}{Dave Cummings}, \bibinfo{person}{Matthias Plappert}, \bibinfo{person}{Fotios Chantzis}, \bibinfo{person}{Elizabeth Barnes}, \bibinfo{person}{Ariel Herbert-Voss}, \bibinfo{person}{William~Hebgen Guss}, \bibinfo{person}{Alex Nichol}, \bibinfo{person}{Alex Paino}, \bibinfo{person}{Nikolas Tezak}, \bibinfo{person}{Jie Tang}, \bibinfo{person}{Igor Babuschkin}, \bibinfo{person}{Suchir Balaji}, \bibinfo{person}{Shantanu Jain}, \bibinfo{person}{William Saunders}, \bibinfo{person}{Christopher Hesse}, \bibinfo{person}{Andrew~N. Carr}, \bibinfo{person}{Jan Leike}, \bibinfo{person}{Josh Achiam}, \bibinfo{person}{Vedant Misra}, \bibinfo{person}{Evan Morikawa}, \bibinfo{person}{Alec Radford}, \bibinfo{person}{Matthew Knight}, \bibinfo{person}{Miles Brundage}, \bibinfo{person}{Mira Murati}, \bibinfo{person}{Katie Mayer}, \bibinfo{person}{Peter Welinder}, \bibinfo{person}{Bob McGrew}, \bibinfo{person}{Dario Amodei}, \bibinfo{person}{Sam McCandlish}, \bibinfo{person}{Ilya Sutskever}, {and} \bibinfo{person}{Wojciech Zaremba}.} \bibinfo{year}{2021}\natexlab{}.
\newblock \bibinfo{title}{Evaluating {Large} {Language} {Models} {Trained} on {Code}}.
\newblock
\newblock
\urldef\tempurl%
\url{http://arxiv.org/abs/2107.03374}
\showURL{%
\tempurl}
\newblock
\shownote{arXiv:2107.03374 [cs]}.


\bibitem[Chen et~al\mbox{.}(2019)]%
        {chen_gmail_2019}
\bibfield{author}{\bibinfo{person}{Mia~Xu Chen}, \bibinfo{person}{Benjamin~N. Lee}, \bibinfo{person}{Gagan Bansal}, \bibinfo{person}{Yuan Cao}, \bibinfo{person}{Shuyuan Zhang}, \bibinfo{person}{Justin Lu}, \bibinfo{person}{Jackie Tsay}, \bibinfo{person}{Yinan Wang}, \bibinfo{person}{Andrew~M. Dai}, \bibinfo{person}{Zhifeng Chen}, \bibinfo{person}{Timothy Sohn}, {and} \bibinfo{person}{Yonghui Wu}.} \bibinfo{year}{2019}\natexlab{}.
\newblock \bibinfo{title}{Gmail {Smart} {Compose}: {Real}-{Time} {Assisted} {Writing}}.
\newblock
\newblock
\urldef\tempurl%
\url{http://arxiv.org/abs/1906.00080}
\showURL{%
\tempurl}
\newblock
\shownote{arXiv:1906.00080 [cs]}.


\bibitem[Chien et~al\mbox{.}(2023)]%
        {chien2023reducing}
\bibfield{author}{\bibinfo{person}{Andrew~A Chien}, \bibinfo{person}{Liuzixuan Lin}, \bibinfo{person}{Hai Nguyen}, \bibinfo{person}{Varsha Rao}, \bibinfo{person}{Tristan Sharma}, {and} \bibinfo{person}{Rajini Wijayawardana}.} \bibinfo{year}{2023}\natexlab{}.
\newblock \showarticletitle{Reducing the Carbon Impact of Generative AI Inference (today and in 2035)}. In \bibinfo{booktitle}{\emph{Proceedings of the 2nd Workshop on Sustainable Computer Systems}}. \bibinfo{pages}{1--7}.
\newblock


\bibitem[Codeium(2023)]%
        {codeium_2023}
Codeium \bibinfo{year}{2023}\natexlab{}.
\newblock \bibinfo{title}{Codeium - Free AI Code Completions}.
\newblock \bibinfo{howpublished}{Online}.
\newblock
\urldef\tempurl%
\url{https://codeium.com/}
\showURL{%
\tempurl}


\bibitem[Evtikhiev et~al\mbox{.}(2023)]%
        {evtikhiev_out_2023}
\bibfield{author}{\bibinfo{person}{Mikhail Evtikhiev}, \bibinfo{person}{Egor Bogomolov}, \bibinfo{person}{Yaroslav Sokolov}, {and} \bibinfo{person}{Timofey Bryksin}.} \bibinfo{year}{2023}\natexlab{}.
\newblock \showarticletitle{Out of the {BLEU}: how should we assess quality of the {Code} {Generation} models?}
\newblock \bibinfo{journal}{\emph{Journal of Systems and Software}}  \bibinfo{volume}{203} (\bibinfo{date}{Sept.} \bibinfo{year}{2023}), \bibinfo{pages}{111741}.
\newblock
\showISSN{01641212}
\urldef\tempurl%
\url{https://doi.org/10.1016/j.jss.2023.111741}
\showDOI{\tempurl}
\newblock
\shownote{arXiv:2208.03133 [cs]}.


\bibitem[Feurer et~al\mbox{.}(2020)]%
        {feurer_auto-sklearn_2020}
\bibfield{author}{\bibinfo{person}{Matthias Feurer}, \bibinfo{person}{Katharina Eggensperger}, \bibinfo{person}{Stefan Falkner}, \bibinfo{person}{Marius Lindauer}, {and} \bibinfo{person}{Frank Hutter}.} \bibinfo{year}{2020}\natexlab{}.
\newblock \showarticletitle{Auto-{Sklearn} 2.0: {Hands}-free {AutoML} via {Meta}-{Learning}}.
\newblock \bibinfo{journal}{\emph{arXiv:2007.04074 [cs.LG]}} (\bibinfo{year}{2020}).
\newblock


\bibitem[Feurer et~al\mbox{.}(2015)]%
        {feurer_efficient_2015}
\bibfield{author}{\bibinfo{person}{Matthias Feurer}, \bibinfo{person}{Aaron Klein}, \bibinfo{person}{Katharina Eggensperger}, \bibinfo{person}{Jost Springenberg}, \bibinfo{person}{Manuel Blum}, {and} \bibinfo{person}{Frank Hutter}.} \bibinfo{year}{2015}\natexlab{}.
\newblock \showarticletitle{Efficient and {Robust} {Automated} {Machine} {Learning}}. In \bibinfo{booktitle}{\emph{Advances in {Neural} {Information} {Processing} {Systems} 28 (2015)}}. \bibinfo{pages}{2962--2970}.
\newblock


\bibitem[Fried et~al\mbox{.}(2023)]%
        {fried_incoder_2023}
\bibfield{author}{\bibinfo{person}{Daniel Fried}, \bibinfo{person}{Armen Aghajanyan}, \bibinfo{person}{Jessy Lin}, \bibinfo{person}{Sida Wang}, \bibinfo{person}{Eric Wallace}, \bibinfo{person}{Freda Shi}, \bibinfo{person}{Ruiqi Zhong}, \bibinfo{person}{Wen-tau Yih}, \bibinfo{person}{Luke Zettlemoyer}, {and} \bibinfo{person}{Mike Lewis}.} \bibinfo{year}{2023}\natexlab{}.
\newblock \bibinfo{title}{{InCoder}: {A} {Generative} {Model} for {Code} {Infilling} and {Synthesis}}.
\newblock
\newblock
\urldef\tempurl%
\url{http://arxiv.org/abs/2204.05999}
\showURL{%
\tempurl}
\newblock
\shownote{arXiv:2204.05999 [cs]}.


\bibitem[Gemini Code Assist(2023)]%
        {gemini_code_assist_2023}
Gemini Code Assist \bibinfo{year}{2023}\natexlab{}.
\newblock \bibinfo{title}{Gemin Code Assist}.
\newblock \bibinfo{howpublished}{Online}.
\newblock
\urldef\tempurl%
\url{https://cloud.google.com/products/gemini/code-assist}
\showURL{%
\tempurl}


\bibitem[GitHub Copilot(2021)]%
        {copilot_2021}
GitHub Copilot \bibinfo{year}{2021}\natexlab{}.
\newblock \bibinfo{title}{GitHub Copilot: Your AI Pair Programmer}.
\newblock \bibinfo{howpublished}{Online}.
\newblock
\urldef\tempurl%
\url{https://github.com/features/copilot}
\showURL{%
\tempurl}


\bibitem[Guo et~al\mbox{.}(2022)]%
        {guo_unixcoder_2022}
\bibfield{author}{\bibinfo{person}{Daya Guo}, \bibinfo{person}{Shuai Lu}, \bibinfo{person}{Nan Duan}, \bibinfo{person}{Yanlin Wang}, \bibinfo{person}{Ming Zhou}, {and} \bibinfo{person}{Jian Yin}.} \bibinfo{year}{2022}\natexlab{}.
\newblock \bibinfo{title}{{UniXcoder}: {Unified} {Cross}-{Modal} {Pre}-training for {Code} {Representation}}.
\newblock
\newblock
\urldef\tempurl%
\url{http://arxiv.org/abs/2203.03850}
\showURL{%
\tempurl}
\newblock
\shownote{arXiv:2203.03850 [cs]}.


\bibitem[Guo et~al\mbox{.}(2024)]%
        {guo_deepseek-coder_2024}
\bibfield{author}{\bibinfo{person}{Daya Guo}, \bibinfo{person}{Qihao Zhu}, \bibinfo{person}{Dejian Yang}, \bibinfo{person}{Zhenda Xie}, \bibinfo{person}{Kai Dong}, \bibinfo{person}{Wentao Zhang}, \bibinfo{person}{Guanting Chen}, \bibinfo{person}{Xiao Bi}, \bibinfo{person}{Y. Wu}, \bibinfo{person}{Y.~K. Li}, \bibinfo{person}{Fuli Luo}, \bibinfo{person}{Yingfei Xiong}, {and} \bibinfo{person}{Wenfeng Liang}.} \bibinfo{year}{2024}\natexlab{}.
\newblock \bibinfo{title}{{DeepSeek}-{Coder}: {When} the {Large} {Language} {Model} {Meets} {Programming} -- {The} {Rise} of {Code} {Intelligence}}.
\newblock
\newblock
\urldef\tempurl%
\url{http://arxiv.org/abs/2401.14196}
\showURL{%
\tempurl}
\newblock
\shownote{arXiv:2401.14196 [cs]}.


\bibitem[Harding and Kloster(2024)]%
        {harding_coding_2024}
\bibfield{author}{\bibinfo{person}{William Harding} {and} \bibinfo{person}{Matthew Kloster}.} \bibinfo{year}{2024}\natexlab{}.
\newblock \bibinfo{booktitle}{\emph{Coding on {Copilot}: 2023 {Data} {Shows} {Downward} {Pressure} on {Code} {Quality}}}.
\newblock \bibinfo{type}{Whitepaper}. \bibinfo{institution}{GitClear}. \bibinfo{pages}{24} pages.
\newblock
\urldef\tempurl%
\url{https://gitclear-public.s3.us-west-2.amazonaws.com/Coding-on-Copilot-2024-Developer-Research.pdf}
\showURL{%
\tempurl}


\bibitem[Huh et~al\mbox{.}(2023)]%
        {huh_low-rank_2023}
\bibfield{author}{\bibinfo{person}{Minyoung Huh}, \bibinfo{person}{Hossein Mobahi}, \bibinfo{person}{Richard Zhang}, \bibinfo{person}{Brian Cheung}, \bibinfo{person}{Pulkit Agrawal}, {and} \bibinfo{person}{Phillip Isola}.} \bibinfo{year}{2023}\natexlab{}.
\newblock \bibinfo{title}{The {Low}-{Rank} {Simplicity} {Bias} in {Deep} {Networks}}.
\newblock
\newblock
\urldef\tempurl%
\url{http://arxiv.org/abs/2103.10427}
\showURL{%
\tempurl}
\newblock
\shownote{arXiv:2103.10427 [cs]}.


\bibitem[Izadi et~al\mbox{.}(2022)]%
        {izadi_codefill_2022}
\bibfield{author}{\bibinfo{person}{Maliheh Izadi}, \bibinfo{person}{Roberta Gismondi}, {and} \bibinfo{person}{Georgios Gousios}.} \bibinfo{year}{2022}\natexlab{}.
\newblock \showarticletitle{{CodeFill}: {Multi}-token {Code} {Completion} by {Jointly} {Learning} from {Structure} and {Naming} {Sequences}}. In \bibinfo{booktitle}{\emph{Proceedings of the 44th {International} {Conference} on {Software} {Engineering}}}. \bibinfo{pages}{401--412}.
\newblock
\urldef\tempurl%
\url{https://doi.org/10.1145/3510003.3510172}
\showDOI{\tempurl}
\newblock
\shownote{arXiv:2202.06689 [cs]}.


\bibitem[Izadi et~al\mbox{.}(2024)]%
        {izadi_language_2024}
\bibfield{author}{\bibinfo{person}{Maliheh Izadi}, \bibinfo{person}{Jonathan Katzy}, \bibinfo{person}{Tim van Dam}, \bibinfo{person}{Marc Otten}, \bibinfo{person}{Razvan~Mihai Popescu}, {and} \bibinfo{person}{Arie van Deursen}.} \bibinfo{year}{2024}\natexlab{}.
\newblock \showarticletitle{Language {Models} for {Code} {Completion}: {A} {Practical} {Evaluation}}. In \bibinfo{booktitle}{\emph{46th {International} {Conference} on {Software} {Engineering} ({ICSE})}}. \bibinfo{publisher}{ACM/IEEE}.
\newblock
\urldef\tempurl%
\url{http://arxiv.org/abs/2402.16197}
\showURL{%
\tempurl}
\newblock
\shownote{arXiv:2402.16197 [cs]}.


\bibitem[Jetbrains AI(2023)]%
        {jetbrains_ai_2023}
Jetbrains AI \bibinfo{year}{2023}\natexlab{}.
\newblock \bibinfo{title}{Jetbrains AI Service an In-IDE Assistant}.
\newblock \bibinfo{howpublished}{Online}.
\newblock
\urldef\tempurl%
\url{https://www.jetbrains.com/ai/}
\showURL{%
\tempurl}


\bibitem[Johnson et~al\mbox{.}(2023)]%
        {johnson_r-u-sure_2023}
\bibfield{author}{\bibinfo{person}{Daniel~D. Johnson}, \bibinfo{person}{Daniel Tarlow}, {and} \bibinfo{person}{Christian Walder}.} \bibinfo{year}{2023}\natexlab{}.
\newblock \showarticletitle{R-{U}-{SURE}? {Uncertainty}-{Aware} {Code} {Suggestions} {By} {Maximizing} {Utility} {Across} {Random} {User} {Intents}}.
\newblock  (\bibinfo{year}{2023}).
\newblock
\urldef\tempurl%
\url{https://doi.org/10.48550/ARXIV.2303.00732}
\showDOI{\tempurl}
\newblock
\shownote{Publisher: arXiv Version Number: 2}.


\bibitem[Katzy et~al\mbox{.}(2023)]%
        {katzy_impact_2023}
\bibfield{author}{\bibinfo{person}{Jonathan Katzy}, \bibinfo{person}{Maliheh Izadi}, {and} \bibinfo{person}{Arie van Deursen}.} \bibinfo{year}{2023}\natexlab{}.
\newblock \bibinfo{title}{On the {Impact} of {Language} {Selection} for {Training} and {Evaluating} {Programming} {Language} {Models}}.
\newblock
\newblock
\urldef\tempurl%
\url{http://arxiv.org/abs/2308.13354}
\showURL{%
\tempurl}
\newblock
\shownote{arXiv:2308.13354 [cs]}.


\bibitem[Kwon et~al\mbox{.}(2023)]%
        {kwon_efficient_2023}
\bibfield{author}{\bibinfo{person}{Woosuk Kwon}, \bibinfo{person}{Zhuohan Li}, \bibinfo{person}{Siyuan Zhuang}, \bibinfo{person}{Ying Sheng}, \bibinfo{person}{Lianmin Zheng}, \bibinfo{person}{Cody~Hao Yu}, \bibinfo{person}{Joseph~E. Gonzalez}, \bibinfo{person}{Hao Zhang}, {and} \bibinfo{person}{Ion Stoica}.} \bibinfo{year}{2023}\natexlab{}.
\newblock \bibinfo{title}{Efficient {Memory} {Management} for {Large} {Language} {Model} {Serving} with {PagedAttention}}.
\newblock
\newblock
\urldef\tempurl%
\url{http://arxiv.org/abs/2309.06180}
\showURL{%
\tempurl}
\newblock
\shownote{arXiv:2309.06180 [cs]}.


\bibitem[Liang et~al\mbox{.}(2023)]%
        {liang_understanding_2023}
\bibfield{author}{\bibinfo{person}{Jenny~T. Liang}, \bibinfo{person}{Chenyang Yang}, {and} \bibinfo{person}{Brad~A. Myers}.} \bibinfo{year}{2023}\natexlab{}.
\newblock \bibinfo{title}{Understanding the {Usability} of {AI} {Programming} {Assistants}}.
\newblock
\newblock
\urldef\tempurl%
\url{http://arxiv.org/abs/2303.17125}
\showURL{%
\tempurl}
\newblock
\shownote{arXiv:2303.17125 [cs]}.


\bibitem[Liu et~al\mbox{.}(2019)]%
        {liu_roberta_2019}
\bibfield{author}{\bibinfo{person}{Yinhan Liu}, \bibinfo{person}{Myle Ott}, \bibinfo{person}{Naman Goyal}, \bibinfo{person}{Jingfei Du}, \bibinfo{person}{Mandar Joshi}, \bibinfo{person}{Danqi Chen}, \bibinfo{person}{Omer Levy}, \bibinfo{person}{Mike Lewis}, \bibinfo{person}{Luke Zettlemoyer}, {and} \bibinfo{person}{Veselin Stoyanov}.} \bibinfo{year}{2019}\natexlab{}.
\newblock \bibinfo{title}{{RoBERTa}: {A} {Robustly} {Optimized} {BERT} {Pretraining} {Approach}}.
\newblock
\newblock
\urldef\tempurl%
\url{http://arxiv.org/abs/1907.11692}
\showURL{%
\tempurl}
\newblock
\shownote{arXiv:1907.11692 [cs]}.


\bibitem[Lozhkov et~al\mbox{.}(2024)]%
        {lozhkov_starcoder_2024}
\bibfield{author}{\bibinfo{person}{Anton Lozhkov}, \bibinfo{person}{Raymond Li}, \bibinfo{person}{Loubna~Ben Allal}, \bibinfo{person}{Federico Cassano}, \bibinfo{person}{Joel Lamy-Poirier}, \bibinfo{person}{Nouamane Tazi}, \bibinfo{person}{Ao Tang}, \bibinfo{person}{Dmytro Pykhtar}, \bibinfo{person}{Jiawei Liu}, \bibinfo{person}{Yuxiang Wei}, \bibinfo{person}{Tianyang Liu}, \bibinfo{person}{Max Tian}, \bibinfo{person}{Denis Kocetkov}, \bibinfo{person}{Arthur Zucker}, \bibinfo{person}{Younes Belkada}, \bibinfo{person}{Zijian Wang}, \bibinfo{person}{Qian Liu}, \bibinfo{person}{Dmitry Abulkhanov}, \bibinfo{person}{Indraneil Paul}, \bibinfo{person}{Zhuang Li}, \bibinfo{person}{Wen-Ding Li}, \bibinfo{person}{Megan Risdal}, \bibinfo{person}{Jia Li}, \bibinfo{person}{Jian Zhu}, \bibinfo{person}{Terry~Yue Zhuo}, \bibinfo{person}{Evgenii Zheltonozhskii}, \bibinfo{person}{Nii Osae~Osae Dade}, \bibinfo{person}{Wenhao Yu}, \bibinfo{person}{Lucas Krauß}, \bibinfo{person}{Naman Jain}, \bibinfo{person}{Yixuan Su}, \bibinfo{person}{Xuanli He}, \bibinfo{person}{Manan Dey}, \bibinfo{person}{Edoardo Abati}, \bibinfo{person}{Yekun Chai}, \bibinfo{person}{Niklas Muennighoff}, \bibinfo{person}{Xiangru Tang}, \bibinfo{person}{Muhtasham Oblokulov}, \bibinfo{person}{Christopher Akiki}, \bibinfo{person}{Marc Marone}, \bibinfo{person}{Chenghao Mou}, \bibinfo{person}{Mayank Mishra}, \bibinfo{person}{Alex Gu}, \bibinfo{person}{Binyuan Hui}, \bibinfo{person}{Tri Dao}, \bibinfo{person}{Armel Zebaze}, \bibinfo{person}{Olivier Dehaene}, \bibinfo{person}{Nicolas Patry}, \bibinfo{person}{Canwen Xu}, \bibinfo{person}{Julian McAuley}, \bibinfo{person}{Han Hu}, \bibinfo{person}{Torsten Scholak}, \bibinfo{person}{Sebastien Paquet}, \bibinfo{person}{Jennifer Robinson}, \bibinfo{person}{Carolyn~Jane Anderson}, \bibinfo{person}{Nicolas Chapados}, \bibinfo{person}{Mostofa Patwary}, \bibinfo{person}{Nima Tajbakhsh}, \bibinfo{person}{Yacine Jernite}, \bibinfo{person}{Carlos~Muñoz Ferrandis}, \bibinfo{person}{Lingming Zhang}, \bibinfo{person}{Sean Hughes}, \bibinfo{person}{Thomas Wolf}, \bibinfo{person}{Arjun Guha}, \bibinfo{person}{Leandro von Werra}, {and} \bibinfo{person}{Harm de Vries}.} \bibinfo{year}{2024}\natexlab{}.
\newblock \bibinfo{title}{{StarCoder} 2 and {The} {Stack} v2: {The} {Next} {Generation}}.
\newblock
\newblock
\urldef\tempurl%
\url{http://arxiv.org/abs/2402.19173}
\showURL{%
\tempurl}
\newblock
\shownote{arXiv:2402.19173 [cs]}.


\bibitem[Lu et~al\mbox{.}(2022)]%
        {lu_reacc_2022}
\bibfield{author}{\bibinfo{person}{Shuai Lu}, \bibinfo{person}{Nan Duan}, \bibinfo{person}{Hojae Han}, \bibinfo{person}{Daya Guo}, \bibinfo{person}{Seung-won Hwang}, {and} \bibinfo{person}{Alexey Svyatkovskiy}.} \bibinfo{year}{2022}\natexlab{}.
\newblock \bibinfo{title}{{ReACC}: {A} {Retrieval}-{Augmented} {Code} {Completion} {Framework}}.
\newblock
\newblock
\urldef\tempurl%
\url{http://arxiv.org/abs/2203.07722}
\showURL{%
\tempurl}
\newblock
\shownote{arXiv:2203.07722 [cs]}.


\bibitem[Lu et~al\mbox{.}(2021)]%
        {lu_codexglue_2021}
\bibfield{author}{\bibinfo{person}{Shuai Lu}, \bibinfo{person}{Daya Guo}, \bibinfo{person}{Shuo Ren}, \bibinfo{person}{Junjie Huang}, \bibinfo{person}{Alexey Svyatkovskiy}, \bibinfo{person}{Ambrosio Blanco}, \bibinfo{person}{Colin Clement}, \bibinfo{person}{Dawn Drain}, \bibinfo{person}{Daxin Jiang}, \bibinfo{person}{Duyu Tang}, \bibinfo{person}{Ge Li}, \bibinfo{person}{Lidong Zhou}, \bibinfo{person}{Linjun Shou}, \bibinfo{person}{Long Zhou}, \bibinfo{person}{Michele Tufano}, \bibinfo{person}{Ming Gong}, \bibinfo{person}{Ming Zhou}, \bibinfo{person}{Nan Duan}, \bibinfo{person}{Neel Sundaresan}, \bibinfo{person}{Shao~Kun Deng}, \bibinfo{person}{Shengyu Fu}, {and} \bibinfo{person}{Shujie Liu}.} \bibinfo{year}{2021}\natexlab{}.
\newblock \bibinfo{title}{{CodeXGLUE}: {A} {Machine} {Learning} {Benchmark} {Dataset} for {Code} {Understanding} and {Generation}}.
\newblock
\newblock
\urldef\tempurl%
\url{http://arxiv.org/abs/2102.04664}
\showURL{%
\tempurl}
\newblock
\shownote{arXiv:2102.04664 [cs]}.


\bibitem[McKinzie et~al\mbox{.}(2024)]%
        {mckinzie_mm1_2024}
\bibfield{author}{\bibinfo{person}{Brandon McKinzie}, \bibinfo{person}{Zhe Gan}, \bibinfo{person}{Jean-Philippe Fauconnier}, \bibinfo{person}{Sam Dodge}, \bibinfo{person}{Bowen Zhang}, \bibinfo{person}{Philipp Dufter}, \bibinfo{person}{Dhruti Shah}, \bibinfo{person}{Xianzhi Du}, \bibinfo{person}{Futang Peng}, \bibinfo{person}{Floris Weers}, \bibinfo{person}{Anton Belyi}, \bibinfo{person}{Haotian Zhang}, \bibinfo{person}{Karanjeet Singh}, \bibinfo{person}{Doug Kang}, \bibinfo{person}{Ankur Jain}, \bibinfo{person}{Hongyu Hè}, \bibinfo{person}{Max Schwarzer}, \bibinfo{person}{Tom Gunter}, \bibinfo{person}{Xiang Kong}, \bibinfo{person}{Aonan Zhang}, \bibinfo{person}{Jianyu Wang}, \bibinfo{person}{Chong Wang}, \bibinfo{person}{Nan Du}, \bibinfo{person}{Tao Lei}, \bibinfo{person}{Sam Wiseman}, \bibinfo{person}{Mark Lee}, \bibinfo{person}{Zirui Wang}, \bibinfo{person}{Ruoming Pang}, \bibinfo{person}{Peter Grasch}, \bibinfo{person}{Alexander Toshev}, {and} \bibinfo{person}{Yinfei Yang}.} \bibinfo{year}{2024}\natexlab{}.
\newblock \bibinfo{title}{{MM1}: {Methods}, {Analysis} \& {Insights} from {Multimodal} {LLM} {Pre}-training}.
\newblock
\newblock
\urldef\tempurl%
\url{http://arxiv.org/abs/2403.09611}
\showURL{%
\tempurl}
\newblock
\shownote{arXiv:2403.09611 [cs]}.


\bibitem[Mok(2024)]%
        {chatgpt_cost}
\bibfield{author}{\bibinfo{person}{Aaron Mok}.} \bibinfo{year}{2024}\natexlab{}.
\newblock \bibinfo{title}{Estimated Cost of ChatGPT}.
\newblock
\newblock
\urldef\tempurl%
\url{https://www.businessinsider.com/how-much-chatgpt-costs-openai-to-run-estimate-report-2023-4?international=true&r=US&IR=T}
\showURL{%
\tempurl}


\bibitem[Mozannar et~al\mbox{.}(2023a)]%
        {mozannar_reading_2023}
\bibfield{author}{\bibinfo{person}{Hussein Mozannar}, \bibinfo{person}{Gagan Bansal}, \bibinfo{person}{Adam Fourney}, {and} \bibinfo{person}{Eric Horvitz}.} \bibinfo{year}{2023}\natexlab{a}.
\newblock \bibinfo{title}{Reading {Between} the {Lines}: {Modeling} {User} {Behavior} and {Costs} in {AI}-{Assisted} {Programming}}.
\newblock
\newblock
\urldef\tempurl%
\url{http://arxiv.org/abs/2210.14306}
\showURL{%
\tempurl}
\newblock
\shownote{arXiv:2210.14306 [cs]}.


\bibitem[Mozannar et~al\mbox{.}(2023b)]%
        {mozannar_when_2023}
\bibfield{author}{\bibinfo{person}{Hussein Mozannar}, \bibinfo{person}{Gagan Bansal}, \bibinfo{person}{Adam Fourney}, {and} \bibinfo{person}{Eric Horvitz}.} \bibinfo{year}{2023}\natexlab{b}.
\newblock \bibinfo{title}{When to {Show} a {Suggestion}? {Integrating} {Human} {Feedback} in {AI}-{Assisted} {Programming}}.
\newblock
\newblock
\urldef\tempurl%
\url{http://arxiv.org/abs/2306.04930}
\showURL{%
\tempurl}
\newblock
\shownote{arXiv:2306.04930 [cs]}.


\bibitem[Peng et~al\mbox{.}(2023)]%
        {peng_impact_2023}
\bibfield{author}{\bibinfo{person}{Sida Peng}, \bibinfo{person}{Eirini Kalliamvakou}, \bibinfo{person}{Peter Cihon}, {and} \bibinfo{person}{Mert Demirer}.} \bibinfo{year}{2023}\natexlab{}.
\newblock \bibinfo{title}{The {Impact} of {AI} on {Developer} {Productivity}: {Evidence} from {GitHub} {Copilot}}.
\newblock
\newblock
\urldef\tempurl%
\url{http://arxiv.org/abs/2302.06590}
\showURL{%
\tempurl}
\newblock
\shownote{arXiv:2302.06590 [cs]}.


\bibitem[Prather et~al\mbox{.}(2023)]%
        {prather_its_2023}
\bibfield{author}{\bibinfo{person}{James Prather}, \bibinfo{person}{Brent~N. Reeves}, \bibinfo{person}{Paul Denny}, \bibinfo{person}{Brett~A. Becker}, \bibinfo{person}{Juho Leinonen}, \bibinfo{person}{Andrew Luxton-Reilly}, \bibinfo{person}{Garrett Powell}, \bibinfo{person}{James Finnie-Ansley}, {and} \bibinfo{person}{Eddie~Antonio Santos}.} \bibinfo{year}{2023}\natexlab{}.
\newblock \showarticletitle{"{It}'s {Weird} {That} it {Knows} {What} {I} {Want}": {Usability} and {Interactions} with {Copilot} for {Novice} {Programmers}}.
\newblock  (\bibinfo{date}{April} \bibinfo{year}{2023}).
\newblock
\urldef\tempurl%
\url{https://doi.org/10.48550/ARXIV.2304.02491}
\showDOI{\tempurl}
\newblock
\shownote{Publisher: arXiv Version Number: 1}.


\bibitem[Rozière et~al\mbox{.}(2023)]%
        {roziere_code_2023}
\bibfield{author}{\bibinfo{person}{Baptiste Rozière}, \bibinfo{person}{Jonas Gehring}, \bibinfo{person}{Fabian Gloeckle}, \bibinfo{person}{Sten Sootla}, \bibinfo{person}{Itai Gat}, \bibinfo{person}{Xiaoqing~Ellen Tan}, \bibinfo{person}{Yossi Adi}, \bibinfo{person}{Jingyu Liu}, \bibinfo{person}{Tal Remez}, \bibinfo{person}{Jérémy Rapin}, \bibinfo{person}{Artyom Kozhevnikov}, \bibinfo{person}{Ivan Evtimov}, \bibinfo{person}{Joanna Bitton}, \bibinfo{person}{Manish Bhatt}, \bibinfo{person}{Cristian~Canton Ferrer}, \bibinfo{person}{Aaron Grattafiori}, \bibinfo{person}{Wenhan Xiong}, \bibinfo{person}{Alexandre Défossez}, \bibinfo{person}{Jade Copet}, \bibinfo{person}{Faisal Azhar}, \bibinfo{person}{Hugo Touvron}, \bibinfo{person}{Louis Martin}, \bibinfo{person}{Nicolas Usunier}, \bibinfo{person}{Thomas Scialom}, {and} \bibinfo{person}{Gabriel Synnaeve}.} \bibinfo{year}{2023}\natexlab{}.
\newblock \bibinfo{title}{Code {Llama}: {Open} {Foundation} {Models} for {Code}}.
\newblock
\newblock
\urldef\tempurl%
\url{http://arxiv.org/abs/2308.12950}
\showURL{%
\tempurl}
\newblock
\shownote{arXiv:2308.12950 [cs]}.


\bibitem[Russo(2023)]%
        {russo_navigating_2023}
\bibfield{author}{\bibinfo{person}{Daniel Russo}.} \bibinfo{year}{2023}\natexlab{}.
\newblock \bibinfo{title}{Navigating the {Complexity} of {Generative} {AI} {Adoption} in {Software} {Engineering}}.
\newblock
\newblock
\urldef\tempurl%
\url{http://arxiv.org/abs/2307.06081}
\showURL{%
\tempurl}
\newblock
\shownote{arXiv:2307.06081 [cs]}.


\bibitem[SourceGraph Cody(2023)]%
        {cody_2023}
SourceGraph Cody \bibinfo{year}{2023}\natexlab{}.
\newblock \bibinfo{title}{Cody - AI Coding Assistant}.
\newblock \bibinfo{howpublished}{Online}.
\newblock
\urldef\tempurl%
\url{https://sourcegraph.com/cody}
\showURL{%
\tempurl}


\bibitem[{Stack Overflow}(2023)]%
        {stack_overflow_stack_2023}
\bibfield{author}{\bibinfo{person}{{Stack Overflow}}.} \bibinfo{year}{2023}\natexlab{}.
\newblock \bibinfo{title}{Stack {Overflow} {Developer} {Survey} 2023}.
\newblock
\newblock
\urldef\tempurl%
\url{https://survey.stackoverflow.co/2023/#section-developer-tools-ai-in-the-development-workflow}
\showURL{%
\tempurl}


\bibitem[Sun et~al\mbox{.}(2023)]%
        {sun_dont_2023}
\bibfield{author}{\bibinfo{person}{Zhensu Sun}, \bibinfo{person}{Xiaoning Du}, \bibinfo{person}{Fu Song}, \bibinfo{person}{Shangwen Wang}, \bibinfo{person}{Mingze Ni}, {and} \bibinfo{person}{Li Li}.} \bibinfo{year}{2023}\natexlab{}.
\newblock \showarticletitle{Don't {Complete} {It}! {Preventing} {Unhelpful} {Code} {Completion} for {Productive} and {Sustainable} {Neural} {Code} {Completion} {Systems}}. In \bibinfo{booktitle}{\emph{2023 {IEEE}/{ACM} 45th {International} {Conference} on {Software} {Engineering}: {Companion} {Proceedings} ({ICSE}-{Companion})}}. \bibinfo{publisher}{IEEE}, \bibinfo{address}{Melbourne, Australia}, \bibinfo{pages}{324--325}.
\newblock
\showISBNx{9798350322637}
\urldef\tempurl%
\url{https://doi.org/10.1109/ICSE-Companion58688.2023.00089}
\showDOI{\tempurl}


\bibitem[Tabnine(2023)]%
        {tabnine_2023}
Tabnine \bibinfo{year}{2023}\natexlab{}.
\newblock \bibinfo{title}{Tabnine AI Coding Assistant}.
\newblock \bibinfo{howpublished}{Online}.
\newblock
\urldef\tempurl%
\url{https://www.tabnine.com/}
\showURL{%
\tempurl}


\bibitem[Thakkar(2023)]%
        {thakkar_copilot_2023}
\bibfield{author}{\bibinfo{person}{Parth Thakkar}.} \bibinfo{year}{2023}\natexlab{}.
\newblock \bibinfo{title}{Copilot {Internals}}.
\newblock
\newblock
\urldef\tempurl%
\url{https://thakkarparth007.github.io/copilot-explorer/posts/copilot-internals}
\showURL{%
\tempurl}
\newblock
\shownote{Publication Title: Copilot-Explorer}.


\bibitem[Vaithilingam et~al\mbox{.}(2022)]%
        {vaithilingam_expectation_2022}
\bibfield{author}{\bibinfo{person}{Priyan Vaithilingam}, \bibinfo{person}{Tianyi Zhang}, {and} \bibinfo{person}{Elena~L. Glassman}.} \bibinfo{year}{2022}\natexlab{}.
\newblock \showarticletitle{Expectation vs. {Experience}: {Evaluating} the {Usability} of {Code} {Generation} {Tools} {Powered} by {Large} {Language} {Models}}. In \bibinfo{booktitle}{\emph{{CHI} {Conference} on {Human} {Factors} in {Computing} {Systems} {Extended} {Abstracts}}}. \bibinfo{publisher}{ACM}, \bibinfo{address}{New Orleans LA USA}, \bibinfo{pages}{1--7}.
\newblock
\showISBNx{978-1-4503-9156-6}
\urldef\tempurl%
\url{https://doi.org/10.1145/3491101.3519665}
\showDOI{\tempurl}


\bibitem[Vaswani et~al\mbox{.}(2023)]%
        {vaswani_attention_2023}
\bibfield{author}{\bibinfo{person}{Ashish Vaswani}, \bibinfo{person}{Noam Shazeer}, \bibinfo{person}{Niki Parmar}, \bibinfo{person}{Jakob Uszkoreit}, \bibinfo{person}{Llion Jones}, \bibinfo{person}{Aidan~N. Gomez}, \bibinfo{person}{Lukasz Kaiser}, {and} \bibinfo{person}{Illia Polosukhin}.} \bibinfo{year}{2023}\natexlab{}.
\newblock \bibinfo{title}{Attention {Is} {All} {You} {Need}}.
\newblock
\newblock
\urldef\tempurl%
\url{http://arxiv.org/abs/1706.03762}
\showURL{%
\tempurl}
\newblock
\shownote{arXiv:1706.03762 [cs]}.


\bibitem[Wang et~al\mbox{.}(2023a)]%
        {wang_investigating_2023}
\bibfield{author}{\bibinfo{person}{Ruotong Wang}, \bibinfo{person}{Ruijia Cheng}, \bibinfo{person}{Denae Ford}, {and} \bibinfo{person}{Thomas Zimmermann}.} \bibinfo{year}{2023}\natexlab{a}.
\newblock \bibinfo{title}{Investigating and {Designing} for {Trust} in {AI}-powered {Code} {Generation} {Tools}}.
\newblock
\newblock
\urldef\tempurl%
\url{http://arxiv.org/abs/2305.11248}
\showURL{%
\tempurl}
\newblock
\shownote{arXiv:2305.11248 [cs]}.


\bibitem[Wang et~al\mbox{.}(2023b)]%
        {wang_codet5_2023}
\bibfield{author}{\bibinfo{person}{Yue Wang}, \bibinfo{person}{Hung Le}, \bibinfo{person}{Akhilesh~Deepak Gotmare}, \bibinfo{person}{Nghi D.~Q. Bui}, \bibinfo{person}{Junnan Li}, {and} \bibinfo{person}{Steven C.~H. Hoi}.} \bibinfo{year}{2023}\natexlab{b}.
\newblock \bibinfo{title}{{CodeT5}+: {Open} {Code} {Large} {Language} {Models} for {Code} {Understanding} and {Generation}}.
\newblock
\newblock
\urldef\tempurl%
\url{http://arxiv.org/abs/2305.07922}
\showURL{%
\tempurl}
\newblock
\shownote{arXiv:2305.07922 [cs]}.


\bibitem[Wu et~al\mbox{.}(2022)]%
        {wu_extreme_2022}
\bibfield{author}{\bibinfo{person}{Xiaoxia Wu}, \bibinfo{person}{Zhewei Yao}, \bibinfo{person}{Minjia Zhang}, \bibinfo{person}{Conglong Li}, {and} \bibinfo{person}{Yuxiong He}.} \bibinfo{year}{2022}\natexlab{}.
\newblock \bibinfo{title}{Extreme {Compression} for {Pre}-trained {Transformers} {Made} {Simple} and {Efficient}}.
\newblock
\newblock
\urldef\tempurl%
\url{https://doi.org/10.48550/arXiv.2206.01859}
\showDOI{\tempurl}
\newblock
\shownote{arXiv:2206.01859 [cs]}.


\bibitem[Zhou et~al\mbox{.}(2023)]%
        {zhou_codebertscore_2023}
\bibfield{author}{\bibinfo{person}{Shuyan Zhou}, \bibinfo{person}{Uri Alon}, \bibinfo{person}{Sumit Agarwal}, {and} \bibinfo{person}{Graham Neubig}.} \bibinfo{year}{2023}\natexlab{}.
\newblock \bibinfo{title}{{CodeBERTScore}: {Evaluating} {Code} {Generation} with {Pretrained} {Models} of {Code}}.
\newblock
\newblock
\urldef\tempurl%
\url{http://arxiv.org/abs/2302.05527}
\showURL{%
\tempurl}
\newblock
\shownote{arXiv:2302.05527 [cs]}.


\bibitem[Zhuo(2024)]%
        {zhuo_ice-score_2024}
\bibfield{author}{\bibinfo{person}{Terry~Yue Zhuo}.} \bibinfo{year}{2024}\natexlab{}.
\newblock \bibinfo{title}{{ICE}-{Score}: {Instructing} {Large} {Language} {Models} to {Evaluate} {Code}}.
\newblock
\newblock
\urldef\tempurl%
\url{http://arxiv.org/abs/2304.14317}
\showURL{%
\tempurl}
\newblock
\shownote{arXiv:2304.14317 [cs]}.


\bibitem[Ziegler et~al\mbox{.}(2022)]%
        {ziegler_productivity_2022}
\bibfield{author}{\bibinfo{person}{Albert Ziegler}, \bibinfo{person}{Eirini Kalliamvakou}, \bibinfo{person}{Shawn Simister}, \bibinfo{person}{Ganesh Sittampalam}, \bibinfo{person}{Alice Li}, \bibinfo{person}{Andrew Rice}, \bibinfo{person}{Devon Rifkin}, {and} \bibinfo{person}{Edward Aftandilian}.} \bibinfo{year}{2022}\natexlab{}.
\newblock \bibinfo{title}{Productivity {Assessment} of {Neural} {Code} {Completion}}.
\newblock
\newblock
\urldef\tempurl%
\url{http://arxiv.org/abs/2205.06537}
\showURL{%
\tempurl}
\newblock
\shownote{arXiv:2205.06537 [cs]}.


\end{thebibliography}

%%%%%%%%%%%%%%%%%%%%%%%%%
% %% If your work has an appendix, this is the place to put it.
\newpage
\cleardoublepage

\appendix

\section{Features used in Copilot's Filter}
\label{app:copilot-features}

\begin{figure}[H]
  \centering
  \includegraphics[width=\linewidth]{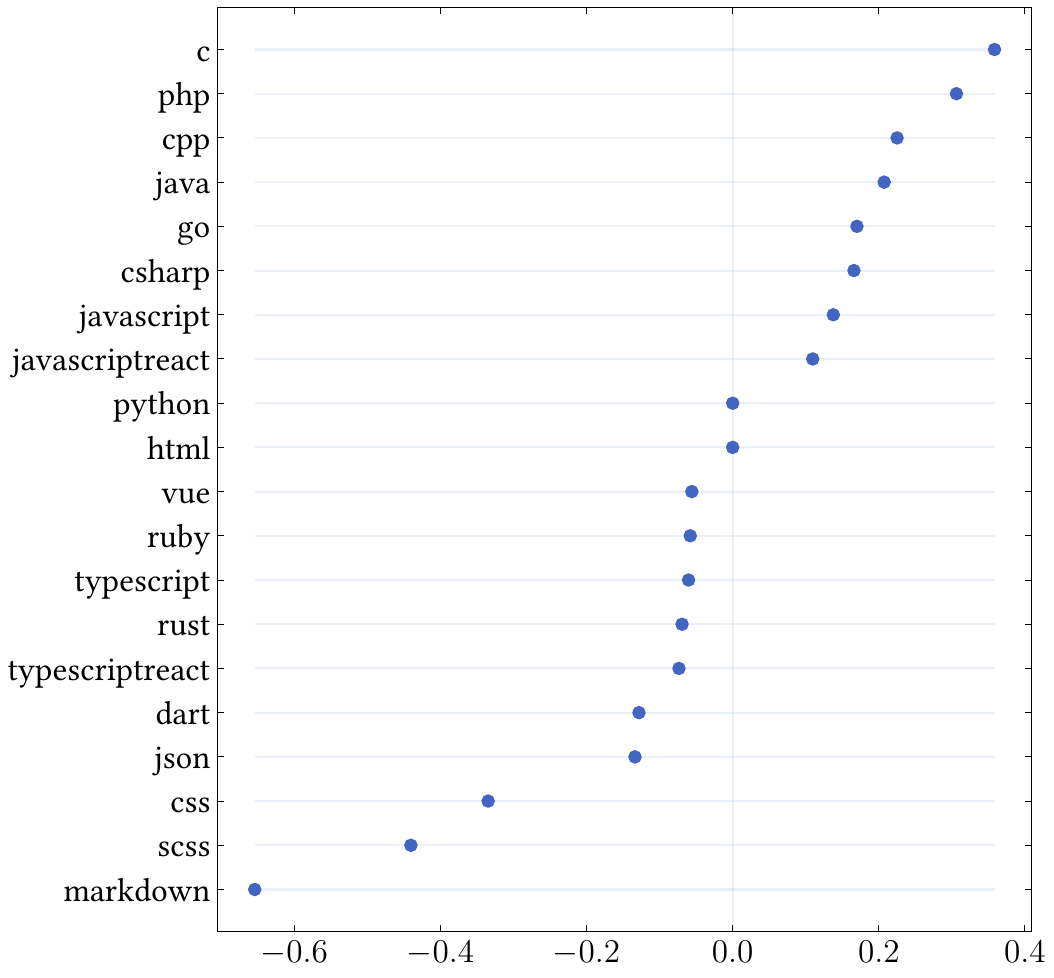}
  \caption{Copilot's Language Map. Higher-scoring languages are more likely to get a completion.}
  \label{fig:copilot-lang-map}
  \Description[
    TODO: desc 
  ]{
    TODO: long desc
  }
\end{figure}

\begin{figure}[H]
  \centering
  \includegraphics[width=\linewidth]{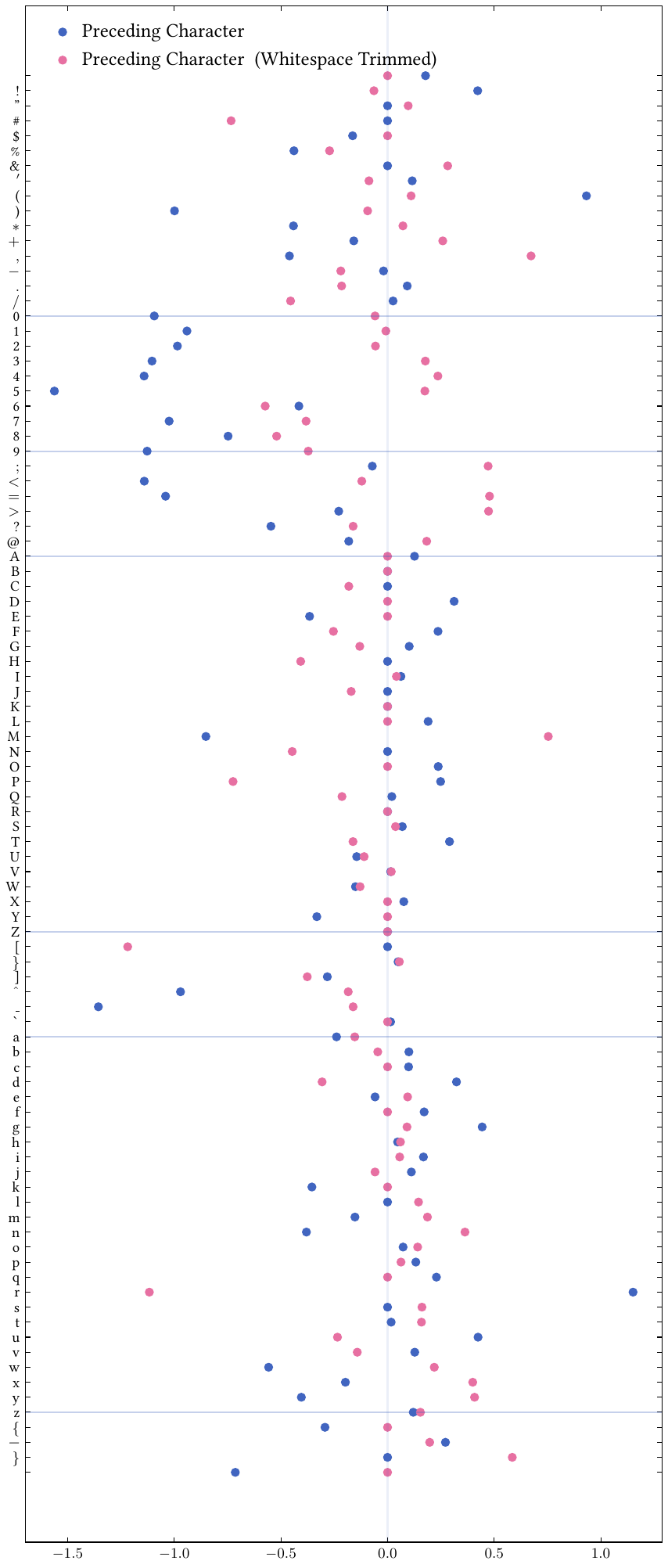}
  \caption{Copilot's Prefix Character Map. Higher-scoring characters (directly before the cursor) are more likely to get a completion.}
  \label{fig:copilot-char-map}
  \Description[
    TODO: desc 
  ]{
    TODO: long desc
  }
\end{figure}

Based on~\cite{thakkar_copilot_2023}, 
we investigated the \texttt{v1.57.7193} (June 2022) version of Copilot 
and list the features used in their logistic classifier in \Cref{tab:copilot-features}. 
The weight is the coefficient in the logistic regression model, 
and the scaling indicates the transformation applied before multiplying with the weight. The language and character maps (last three features) are a one-hot encoding of a fixed set of languages and characters, given in \Cref{fig:copilot-lang-map} and \Cref{fig:copilot-char-map} respectively. 
% TODO: check whether these languages are actually directly taken from Xu et al. 

Before a prompt reaches the filter, two hard-coded rules prevent 
(1) prompts containing fewer than 10 characters, and 
(2) prompts where the cursor is in the middle of a line, by checking whether there is whitespace after the cursor; except if there is a closing character on that line, such as a closing bracket, quote, or semicolon.  

\begin{table}[H]

  \caption{Features Used in Copilot's Filter.} 
  \label{tab:copilot-features}

  \begin{tabular}{lrrr}

    \toprule 
    \textbf{Feature}            & \textbf{Weight}       &\textbf{Input Scaling}   \\
    \midrule                                             
    % data                                               
    Previous filter label       & $0.997 $              & none              \\
    Whitespace after cursor     & $0.700 $              & none              \\
    Time since last label       & $-0.174 $             & log               \\
    Length of last prefix line  & $-0.230 $             & log               \\
    Above, without whitespace   & $0.134 $              & log               \\
    Document length             & $-0.007 $             & log               \\
    Cursor offset               & $0.005 $              & log               \\
    Offset as percentage        & $0.419 $              & none              \\

    % \midrule not allowed according to ACM guidelines
    Document language (map)     & $[-0.654, 0.358]$     & none              \\
    Last prefix char (map)      & $[-1.56, 1.15]$                & none              \\
    Above, without whitespace   & $[-1.12, 0.85]$       & none              \\

    \bottomrule

  \end{tabular}
\end{table}

% Features considered by Xu et al. (2022) vs. Copilot
% C                c
% C#               csharp
% C++              cpp
% Go               go
% Java             java
% JavaScript       javascript, javascriptreact
% PHP              php
% Python           python
% Ruby             ruby
% Rust             rust
% Scala            
% TypeScript       typescript, typescriptreact
%                   html, json
%                   vue 
%                   dart

The languages in \Cref{fig:copilot-lang-map} show that more verbose languages like C and Java are biased to receive more completions, while design-oriented languages like CSS, SCSS, and TypeScript-React are less likely to receive completions. 
Notably, Markdown is least likely to receive completions, likely because the LLMs that generate completions are primarily trained on code; and, it may be especially annoying to developers to receive incorrect completions when they are in a natural-language flow. 

The characters in \Cref{fig:copilot-char-map} reveal some specific usage patterns too. Many letters fall close to 0, indicating that they do not have much influence on the filtering; with the exception of `r', maybe due to the common print-statement completions which tend to be easier to infer. 
Furthermore, if the developer is currently typing a number, this has a negative weight. Presumably, programmers don't want completions while they are writing some uninferrable number. 
Lastly, perhaps surprisingly, the full-stop `.' has near-zero weight, while a comma `,' followed by whitespace has positive weight, implying developers may rely on Copilot to complete list/object structures. 

\section{Comparison of tokenisation Strategies}
\label{app:tokenisation-comparison}

To investigate the effectiveness of our joint prefix-and-suffix tokenisation strategy, 
we compare it to the two alternatives: prefix-only and suffix-only tokenisation.
To this end, we train RoBERTa models from the \texttt{huggingface/CodeBERTa-small-v1} checkpoint
\footnote{https://huggingface.co/huggingface/CodeBERTa-small-v1}. 
We use the same hyperparameters for all models: 
a learning rate of $2 \times 10^{-5}$,
a batch size of 8,
and train for 3 epochs.
% We cross-validate with 5 folds.  
% TODO: STDEV

\begin{table}[H]

  \caption{Comparison of tokenisation Strategies.} 
  \label{tab:tokenisation-comparison}

  \begin{tabular}{llrrrr}

    \toprule 
    % headers 
    \textbf{Tokenisation} 
                 && Manual       & Auto/acc.     & Auto/rej.     & \textbf{Average} \\
                 % & &$n=6118$       &$n=431$        &$n=15889$      &                \\
    \midrule                                            
    
    % data                                             
    Suffix-only  && $89.3$        & $\mathbf{93.0}$ & $24.1$    & $68.8$             \\
    Prefix-only  && $97.6$        & $61.0$        & $\mathbf{74.7}$    & $77.8$      \\
    Joint        && $\mathbf{98.6}$  & $73.5$     & $71.7$       & $\mathbf{81.3}$   \\

    \bottomrule

  \end{tabular}
\end{table}

\section{The Effect of Dataset Distribution on Model Performance}
\label{app:data-comparison}

\begin{figure}[H]
  \centering
  \includegraphics[width=\linewidth]{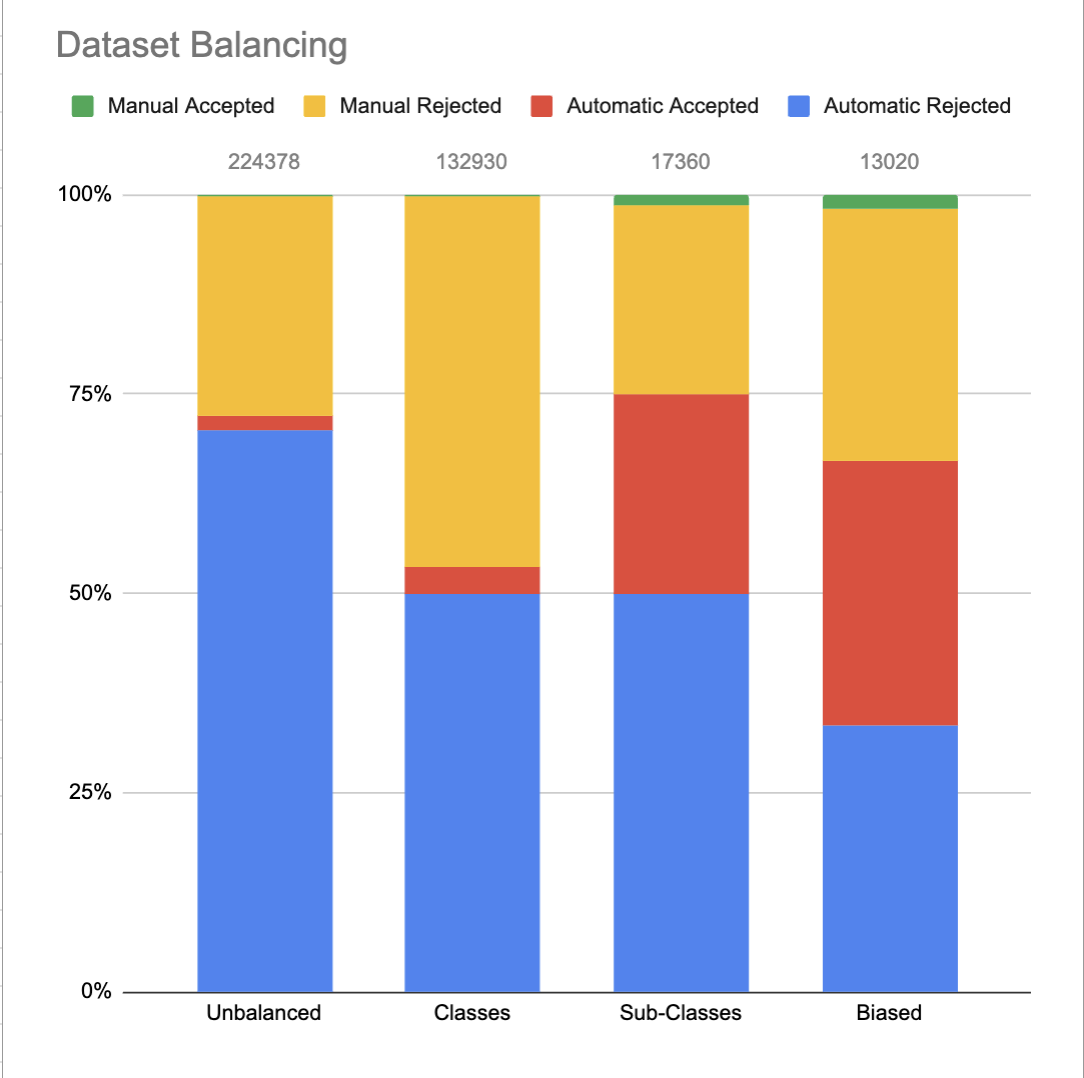}
  \caption{Dataset Distributions}
  \label{fig:dataset-distributions}
  \Description[
    TODO: desc 
  ]{
    TODO: long desc
  }
\end{figure}

Our code completion dataset distribution does not equally represent the invocation types we want to classify, yet they should be weighed roughly equally. 
To remedy this, we experiment with several data distributions (\Cref{fig:dataset-distributions}) for training our models. As shown in \Cref{table:codeberta-data-distribution} for \textsc{CodeBERTa}, and \Cref{table:logres-data-distribution} for the Logistic Regression model, 
find that the biased distribution, giving roughly equal weight to all sub-classes \textit{and} undersampling to better represent manual, accepted invocations, yields the best macro average performance across the sub-classes. 
We observe a similar pattern for all models detailed in \Cref{app:model-comparison}. 
Thus, we use the \textit{biased} distribution as our training/validation dataset. 

For \textsc{CodeBERTa} in \Cref{table:codeberta-data-distribution}, the superior performance on the biased balancing may also be due to our training hyperparameters. We fine-tune for 8 epochs (as opposed to 6 in the rest of this paper), and our models are prone to overfitting due to the small training set size. Future work could perhaps study a data curriculum, going from biased to unbalanced (real-world distribution) to help the model learn classes quickly, and then generalise to data more diverse in textual content.

\begin{table}[h]

  % NOTE: these are not bootstrapped; but don't think it's necessary as the difference is pronounced.
  \caption{\textsc{CodeBERTa} Models under Different Training Data Distributions.} 
  \label{table:codeberta-data-distribution}

  \begin{tabular}{llrrrr}
    \toprule
                    & & Manual        & Auto/acc.     & Auto/rej.     & \textbf{Average}\\
                    % & &$n=6118$       &$n=431$        &$n=15889$      &                 \\
    \midrule                                                                      

    Unbalanced      & & 98.9          & 19.4          & 98.1          & 72.1            \\ 
    Classes         & & 99.2          & 32.2          & 95.7          & 75.7            \\ 
    Sub-Classes     & & 98.2          & 71.6          & 74.4          & 81.3            \\
    \textbf{Biased} & & 98.6          & 78.9          & 67.3          & \textbf{81.6}   \\

    \bottomrule
  \end{tabular}

\end{table}

\begin{table}[h]

  % NOTE: these were bootstrapped
  \caption{Logistic Regression Models (with full telemetry), under Different Training Data Distributions.} 
  \label{table:logres-data-distribution}

  \begin{tabular}{llrrrr}
    \toprule
                    & & Manual        & Auto/acc.     & Auto/rej.     & \textbf{Average} \\
                    % & &$n=6118$       &$n=431$        &$n=15889$      &                \\
    \midrule                                                                      

    Unbalanced      & & 97.4          & 0.0           & 99.8          & 65.8           \\ 
    Classes         & & 97.8          & 0.0           & 99.7          & 65.8           \\ 
    Sub-Classes     & & 97.8          & 35.7          & 88.4          & 74.0           \\
    \textbf{Biased} & & 98.5          & 66.1          & 65.0          & \textbf{76.5}  \\

    \bottomrule
  \end{tabular}

\end{table}

\newpage
\section{\jonberta\textsc{-head} Architecture Experiments}
\label{app:jonberta-head-arch-exp}

We list the 3 \textsc{jonberta-head} variants below, with the 2 options for whether to reinitialise the module or not. 
We hypothesise that the dense variant performs best because dense layers allow the model to learn low-rank embeddings, and speed up training at this scale. 
For comparison, we provide the baseline \textsc{CodeBERTa} model trained on code-context alone.

\begin{table}[h]

  \caption{\textsc{JonBERTa-head} Variants. R~denotes re-initialised modules.} 
  \label{table:jonberta-head-variants}
  \begin{tabular}{lrrrr}
    \toprule
    & Manual      & Auto/acc       & Auto/rej       & \textbf{Avg.} \\
    \midrule
    dense-R       & 98.6 $\pm$ 0.4 & 78.0 $\pm$ 6.2 & 71.4 $\pm$ 5.1 & \textbf{82.7} \\
    proj-R        & 98.6 $\pm$ 0.6 & 76.1 $\pm$ 11.0 & 72.4 $\pm$ 3.6 & 82.3 \\
    dense         & 98.6 $\pm$ 0.5 & 75.6 $\pm$ 6.8 & 72.6 $\pm$ 4.4 & 82.3 \\
    dense-proj-   & 98.6 $\pm$ 0.3 & 73.5 $\pm$ 6.7 & 74.1 $\pm$ 1.8 & 82.1 \\
    dense-proj-R  & 98.5 $\pm$ 0.5 & 72.2 $\pm$ 5.8 & 75.4 $\pm$ 1.9 & 82.0 \\
    proj          & 98.4 $\pm$ 0.5 & 72.9 $\pm$ 7.0 & 74.7 $\pm$ 5.0 & 82.0 \\

    % these are theoretically just fine-tuned RoBERTas
    % --            & 98.5 $\pm$ 0.4 & 74.7 $\pm$ 4.6 & 73.1 $\pm$ 1.2 & 82.1 \\
    % --reinit & 98.5 $\pm$ 0.4 & 74.3 $\pm$ 5.1 & 73.1 $\pm$ 1.2 & 82.0 \\
    % --reinit & 98.2 $\pm$ 0.5 & 74.5 $\pm$ 6.8 & 70.4 $\pm$ 3.4 & 81.0 \\
    % -- & 98.2 $\pm$ 0.5 & 74.4 $\pm$ 8.1 & 70.3 $\pm$ 3.4 & 81.0 \\
    
    \midrule
    \textsc{CodeBERTa}      & 98.5 $\pm$0.4  & 74.7 $\pm$4.6  & 73.1 $\pm$1.2  & 82.1   \\

    \bottomrule

  \end{tabular}
\end{table}

\section{\jonberta\textsc{-attn} Architecture Experiments}
\label{app:jonberta-attn-arch-exp}

It may seem natural to consider telemetry data as simply another modality that can be fed to the transformer akin to multimodal large language models 
\cite{mckinzie_mm1_2024}. 
We choose not to explore this approach as it introduces too many new parameters for the model to learn in our limited-data setting. Furthermore, as far as we know, there are no studies investigating whether an existing natural-language transformer can be extended with additional modalities without reinitialising (all) its weights. To this end, we try learning an embedding for each telemetry feature as a function of that feature, and extend the attention module to attend to these embeddings as well. 

As shown in \Cref{table:jonberta-attn-v1}, we train a variety of layer configurations, from individual layers to triplets of layers. 
We initialise the model from \textsc{CodeBERTa-small-v1}, provided by HuggingFace\footnote{https://huggingface.co/huggingface/CodeBERTa-small-v1}, and fine-tune for 6 epochs on our collected code-completion dataset. 
We note that the final performance of all models is lower than that of \textsc{CodeBERTa}, and hypothesise this may be due to the dual training objective; i.e. we are fine-tuning both as a code-context classification task, and learning telemetry embeddings simultaneously. 

To further investigate, we first fine-tune the \textsc{CodeBERTa} model on code-completion classification for only 3 epochs, and then fine-tune an additional 3 epochs with the additional telemetry features. The results are shown in \Cref{table:jonberta-attn-variants-12}, where the worst-performing model attains classification accuracy on par with the best-performing model from \Cref{table:jonberta-attn-v1}, indicating that better performance is achievable in this training paradigm.

\begin{table}[H]

  \caption{Performance of \textsc{JonBERTa-attn} Variants initialised from \textsc{RoBERTa-small-v1}.} 
  \label{table:jonberta-attn-v1}
  \begin{tabular}{lrrrr}
    \toprule
    Layers    & Manual      & Auto/acc       & Auto/rej       & \textbf{Avg.} \\
    \midrule
    $1, 4, 5$ & 98.4 $\pm$ 0.4 & 75.9 $\pm$ 6.4 & 69.9 $\pm$ 5.8 & \textbf{81.4} \\
    $2, 4, 5$ & 98.4 $\pm$ 0.5 & 77.1 $\pm$ 6.9 & 68.5 $\pm$ 6.4 & 81.3 \\
    $0, 4, 5$ & 98.2 $\pm$ 0.7 & 76.4 $\pm$ 8.0 & 69.1 $\pm$ 5.0 & 81.3 \\
    $0, 1, 4$ & 98.3 $\pm$ 0.4 & 77.3 $\pm$ 5.8 & 68.1 $\pm$ 2.7 & 81.2 \\
    $0, 1, 5$ & 98.3 $\pm$ 0.4 & 77.3 $\pm$ 5.0 & 68.1 $\pm$ 2.8 & 81.2 \\
    $0, 2, 4$ & 98.3 $\pm$ 0.7 & 77.5 $\pm$ 7.1 & 67.9 $\pm$ 5.6 & 81.2 \\
    $1, 2, 3$ & 98.2 $\pm$ 0.5 & 77.6 $\pm$ 5.0 & 67.9 $\pm$ 2.1 & 81.2 \\
    $0, 1, 3$ & 98.4 $\pm$ 0.4 & 78.0 $\pm$ 6.1 & 67.2 $\pm$ 6.1 & 81.2 \\
    $3, 4, 5$ & 98.4 $\pm$ 0.5 & 76.3 $\pm$ 6.2 & 68.7 $\pm$ 3.3 & 81.1 \\
    $0, 2, 5$ & 98.3 $\pm$ 0.4 & 77.4 $\pm$ 5.8 & 67.6 $\pm$ 3.1 & 81.1 \\
    $1, 2, 4$ & 98.4 $\pm$ 0.5 & 77.3 $\pm$ 7.0 & 67.6 $\pm$ 5.0 & 81.1 \\
    $0, 1, 2$ & 98.5 $\pm$ 0.6 & 80.6 $\pm$ 5.2 & 64.1 $\pm$ 6.2 & 81.1 \\
    $0, 3, 5$ & 98.3 $\pm$ 0.5 & 76.0 $\pm$ 7.5 & 68.7 $\pm$ 6.4 & 81.0 \\
    $2, 3, 4$ & 98.3 $\pm$ 0.5 & 76.2 $\pm$ 6.7 & 68.4 $\pm$ 8.2 & 81.0 \\
    $2      $ & 98.0 $\pm$ 0.6 & 77.1 $\pm$ 5.7 & 67.4 $\pm$ 4.0 & 80.8 \\
    $2, 3, 5$ & 98.3 $\pm$ 0.5 & 77.5 $\pm$ 7.2 & 66.6 $\pm$ 5.2 & 80.8 \\
    $0, 3, 4$ & 98.4 $\pm$ 0.5 & 76.6 $\pm$ 5.9 & 67.2 $\pm$ 4.3 & 80.7 \\
    $0, 2, 3$ & 98.3 $\pm$ 0.7 & 74.9 $\pm$ 10.6 & 68.8 $\pm$ 5.9 & 80.7 \\
    $1, 2, 5$ & 98.3 $\pm$ 0.3 & 76.8 $\pm$ 6.3 & 66.8 $\pm$ 6.1 & 80.6 \\
    $1      $ & 98.2 $\pm$ 0.5 & 76.6 $\pm$ 7.7 & 66.9 $\pm$ 3.4 & 80.5 \\
    $0      $ & 98.1 $\pm$ 0.7 & 76.5 $\pm$ 10.0 & 66.8 $\pm$ 5.8 & 80.5 \\
    $1, 3, 5$ & 98.2 $\pm$ 0.6 & 74.7 $\pm$ 10.2 & 68.3 $\pm$ 7.2 & 80.4 \\
    $3      $ & 98.1 $\pm$ 0.8 & 77.5 $\pm$ 7.8 & 65.3 $\pm$ 5.7 & 80.3 \\
    $1, 3, 4$ & 98.3 $\pm$ 0.6 & 74.1 $\pm$ 7.6 & 68.5 $\pm$ 6.6 & 80.3 \\
    $4      $ & 98.2 $\pm$ 0.4 & 77.7 $\pm$ 6.4 & 64.5 $\pm$ 5.3 & 80.1 \\
    $5      $ & 98.2 $\pm$ 0.5 & 78.7 $\pm$ 8.2 & 63.4 $\pm$ 5.9 & 80.1 \\

    \midrule
    \textsc{CodeBERTa}      & 98.5 $\pm$0.4  & 74.7 $\pm$4.6  & 73.1 $\pm$1.2  & \textbf{82.1}   \\

    \bottomrule

  \end{tabular}
\end{table}

\begin{table}[H]

  \caption{\textsc{JonBERTa-attn} variants initialised from our fine-tuned \textsc{CodeBERTa}.} 
  \label{table:jonberta-attn-variants-12}
  \begin{tabular}{lllll}
    \toprule
    Layers & Manual      & Auto/acc       & Auto/rej       & \textbf{Avg.} \\
    \midrule

    $4, 5  $ & 98.6 $\pm$ 0.5 & 76.6 $\pm$ 5.7 & 72.1 $\pm$ 4.2 & 82.4 \\
    $2, 3, 4$ & 98.5 $\pm$ 0.4 & 74.8 $\pm$ 5.9 & 73.8 $\pm$ 2.4 & 82.4 \\
    $0, 1, 4$ & 98.5 $\pm$ 0.4 & 75.2 $\pm$ 7.6 & 73.1 $\pm$ 4.1 & 82.3 \\
    $0, 1, 3$ & 98.5 $\pm$ 0.4 & 74.8 $\pm$ 4.6 & 73.4 $\pm$ 2.0 & 82.3 \\
    $2, 5  $ & 98.5 $\pm$ 0.4 & 74.9 $\pm$ 4.6 & 73.2 $\pm$ 3.0 & 82.2 \\
    $0, 5  $ & 98.6 $\pm$ 0.3 & 74.6 $\pm$ 7.5 & 73.3 $\pm$ 6.1 & 82.2 \\
    $1, 2, 5$ & 98.6 $\pm$ 0.3 & 75.5 $\pm$ 7.2 & 72.4 $\pm$ 2.7 & 82.2 \\
    $2     $ & 98.6 $\pm$ 0.4 & 75.8 $\pm$ 5.9 & 71.9 $\pm$ 3.1 & 82.1 \\
    $3, 4  $ & 98.6 $\pm$ 0.4 & 75.7 $\pm$ 5.3 & 72.0 $\pm$ 2.5 & 82.1 \\
    $3, 4, 5$ & 98.4 $\pm$ 0.4 & 74.4 $\pm$ 5.0 & 73.5 $\pm$ 2.3 & 82.1 \\
    $3, 5  $ & 98.5 $\pm$ 0.3 & 74.2 $\pm$ 6.3 & 73.6 $\pm$ 2.7 & 82.1 \\
    $0, 4, 5$ & 98.5 $\pm$ 0.4 & 75.0 $\pm$ 4.9 & 72.8 $\pm$ 2.3 & 82.1 \\
    $0, 2  $ & 98.7 $\pm$ 0.3 & 77.5 $\pm$ 4.6 & 70.1 $\pm$ 1.9 & 82.1 \\
    $0, 3, 5$ & 98.6 $\pm$ 0.4 & 75.4 $\pm$ 6.2 & 72.3 $\pm$ 2.7 & 82.1 \\
    $2, 3, 5$ & 98.4 $\pm$ 0.4 & 72.4 $\pm$ 6.2 & 75.4 $\pm$ 3.4 & 82.1 \\
    $0, 2, 5$ & 98.5 $\pm$ 0.4 & 74.1 $\pm$ 4.3 & 73.5 $\pm$ 1.5 & 82.1 \\
    $0, 1  $ & 98.7 $\pm$ 0.4 & 75.4 $\pm$ 5.6 & 72.1 $\pm$ 2.1 & 82.0 \\
    $0, 3  $ & 98.7 $\pm$ 0.3 & 76.7 $\pm$ 5.0 & 70.7 $\pm$ 3.0 & 82.0 \\
    $0     $ & 98.6 $\pm$ 0.4 & 75.0 $\pm$ 6.8 & 72.5 $\pm$ 3.5 & 82.0 \\
    $1, 2, 3$ & 98.6 $\pm$ 0.4 & 74.4 $\pm$ 4.4 & 73.1 $\pm$ 1.3 & 82.0 \\
    $1, 2, 4$ & 98.5 $\pm$ 0.4 & 73.1 $\pm$ 7.8 & 74.4 $\pm$ 2.9 & 82.0 \\
    $1, 5  $ & 98.6 $\pm$ 0.6 & 75.3 $\pm$ 9.1 & 72.1 $\pm$ 4.3 & 82.0 \\
    $1, 4  $ & 98.6 $\pm$ 0.5 & 75.8 $\pm$ 7.1 & 71.6 $\pm$ 3.0 & 82.0 \\
    $0, 1, 2$ & 98.6 $\pm$ 0.5 & 74.3 $\pm$ 6.5 & 73.1 $\pm$ 6.0 & 82.0 \\
    $0, 2, 4$ & 98.5 $\pm$ 0.4 & 73.6 $\pm$ 4.6 & 73.7 $\pm$ 3.5 & 81.9 \\
    $0, 2, 3$ & 98.5 $\pm$ 0.5 & 74.3 $\pm$ 5.4 & 73.0 $\pm$ 5.1 & 81.9 \\
    $0, 4  $ & 98.5 $\pm$ 0.4 & 75.0 $\pm$ 5.3 & 72.2 $\pm$ 1.4 & 81.9 \\
    $0, 3, 4$ & 98.6 $\pm$ 0.4 & 76.6 $\pm$ 7.9 & 70.5 $\pm$ 4.2 & 81.9 \\
    $1, 2  $ & 98.6 $\pm$ 0.4 & 74.3 $\pm$ 8.4 & 72.6 $\pm$ 4.2 & 81.9 \\
    $1, 3, 4$ & 98.5 $\pm$ 0.4 & 74.6 $\pm$ 6.8 & 72.6 $\pm$ 3.4 & 81.9 \\
    $0, 1, 5$ & 98.6 $\pm$ 0.5 & 76.4 $\pm$ 10.1 & 70.6 $\pm$ 4.9 & 81.9 \\
    $1, 4, 5$ & 98.6 $\pm$ 0.4 & 73.7 $\pm$ 6.5 & 73.1 $\pm$ 2.2 & 81.8 \\
    $1     $ & 98.5 $\pm$ 0.4 & 73.8 $\pm$ 7.3 & 73.0 $\pm$ 7.5 & 81.8 \\
    $2, 3  $ & 98.5 $\pm$ 0.5 & 72.4 $\pm$ 7.6 & 74.4 $\pm$ 2.9 & 81.8 \\
    $1, 3, 5$ & 98.6 $\pm$ 0.5 & 73.0 $\pm$ 8.1 & 73.7 $\pm$ 4.8 & 81.7 \\
    $2, 4, 5$ & 98.5 $\pm$ 0.5 & 73.9 $\pm$ 5.5 & 72.8 $\pm$ 4.9 & 81.7 \\
    $4     $ & 98.5 $\pm$ 0.5 & 75.5 $\pm$ 9.9 & 71.0 $\pm$ 4.4 & 81.7 \\
    $2, 4  $ & 98.5 $\pm$ 0.4 & 73.7 $\pm$ 6.5 & 72.9 $\pm$ 6.7 & 81.7 \\
    $3     $ & 98.5 $\pm$ 0.3 & 75.0 $\pm$ 8.6 & 71.4 $\pm$ 6.5 & 81.6 \\
    $5     $ & 98.4 $\pm$ 0.5 & 74.6 $\pm$ 6.9 & 71.8 $\pm$ 5.6 & 81.6 \\
    $1, 3  $ & 98.5 $\pm$ 0.6 & 73.8 $\pm$ 7.3 & 71.9 $\pm$ 5.8 & 81.4 \\

    \bottomrule
  \end{tabular}
\end{table}

\section{Alternative Models}
\label{app:model-comparison}
We have exhaustively explored alternative approaches to our completion classification problem, which have guided and motivated our transformer-based approach. 
To this end, we train logistic-regression models on telemetry data, and extend them with various methods of incorporating code context; as well as AutoML models \cite{feurer_efficient_2015,feurer_auto-sklearn_2020}.
Our main finding, as stated in the paper, is that code context is significantly more important for classification accuracy than what can be retrieved from telemetry data alone. 

The alternative models are shown in \Cref{tab:alternative-models}. 
We compare by encoding $n$ prefix and $n$ suffix words/tokens/lines surrounding the cursor. 
It is not fully fair to compare every method with this granularity, as some methods may perform better with e.g. 3 prefix and the entire suffix encoded, but we do so for consistency. 
If we had any remarkable results, this would not be in the appendix in the first place. 

Autosklearn and AutoSklearn2 are ensembles consisting of 22 and 3 models respectively. We omit these models from the body of our study as we prefer simpler solutions over complex ones. We further want to highlight the potential of integrating pre-trained transformers' semantical understanding and contextual capabilities with additional modalities like telemetry feature data. 

\begin{table}[H]

  \caption{Filter accuracy for Logistic Regression classification models with different code-context encodings, given per invocation sub-class. Macro denotes the macro average of the three sub-classes.}
  \label{tab:alternative-models}

  \begin{tabular}{lrrrr}
    \toprule
    Logistic Regr. w/ & Manual       & Auto/acc.     & Auto/rej.     & \textbf{Macro} \\
    \midrule

    % <2.5ms latency per sample. Tokens are one-hot individually!
    % TODO: which tokeniser again? 
    \multicolumn{5}{l}
    {One-Hot Tokens ($\dim = 50821$)}   \\
    3 Tokens          & 99.0          & 77.7          & 54.0          & 77.1        \\
    2 Tokens          & 99.1          & 77.5          & 53.0          & 76.7        \\ 
    1 Token           & 99.1          & 75.2          & 50.7          & 73.9        \\ \\

    % <1ms latency per sample. Tokens are encoded individually, and then concatenated!
    % TODO: add tokeniser
    \multicolumn{5}{l}
    {Tok2Vec ($\dim = 100$)}    \\
    3 tokens          & 97.7          & 74.0          & 58.8          & 76.0        \\ 
    2 tokens          & 98.3          & 75.4          & 56.2          & 76.3        \\ 
    1 token           & 98.5          & 76.3          & 52.7          & 75.5        \\ \\ 

    % 0ms (no tokenisation step). Words are encoded individually, and then concatenated!
    % TODO: add word2vec model used 
    \multicolumn{5}{l}
    {Word2Vec ($\dim = 100$)}                                                        \\
    3 words           & 74.9          & 60.3          & 58.2           & 63.7        \\
    2 words           & 77.5          & 60.6          & 56.3           & 64.2        \\ 
    1 words           & 85.1          & 63.3          & 51.7           & 66.7        \\ \\ 

    % <1ms latency per sample (GPU)
    % Prefix and suffix are embedded either concatenated, or individually and then their embds concatenated!
    \multicolumn{5}{l}
    {SetFit ($\dim = 768$) }\\ 
    1 line            & 78.6          & 58.3          & 66.9            & 70.7        \\ 
    2 lines           & 67.4          & 57.6          & 62.1            & 61.4        \\ 
    3 lines           & 61.3          & 57.3          & 60.8            & 59.9        \\ \\

    % <1ms latency per sample (CPU)
    \multicolumn{5}{l}
    {TF.IDF ($\dim \in [15000, 18500]$)}                                              \\ 
    3 tokens          & 93.0          & 84.5          & 41.6            & 72.9        \\
    2 tokens          & 95.0          & 84.0          & 41.4            & 72.4        \\
    1 token           & 97.0          & 80.5          & 38.4            & 69.4        \\ 
    1 line            & 89.6          & 91.9          & 24.0            & 69.0        \\ \\ 

    % TODO: move these into their own table, as they are not logistic regression but ensembles. 
    AutoSklearn~\cite{feurer_efficient_2015}
                      & 98.9          & 75.9          & 64.6            & 79.4        \\ 
    AutoSklearn2~\cite{feurer_auto-sklearn_2020}
                      & 98.7          & 73.5          & 65.7            & 78.9        \\
   
    \bottomrule
  \end{tabular}

\end{table}

For our One-Hot Tokens and Tok2Vec classifiers, we use the TikToken \texttt{p50k-base} tokeniser from OpenAI's Codex models\footnote{https://github.com/openai/tiktoken}. 
For SetFit, we experiment with three embedding models, and only report \texttt{all-mpnet-base-v2}\footnote{https://huggingface.co/sentence-transformers/all-mpnet-base-v2} here as it marginally outperforms.

\end{document}